\documentclass[conference]{IEEEtran}
\IEEEoverridecommandlockouts
\usepackage{cite}
\usepackage{amsmath,amssymb,amsfonts}
\usepackage[ruled,linesnumbered,vlined]{algorithm2e}
\usepackage{graphicx}
\usepackage{graphicx, subfigure}
\usepackage{float}
\usepackage{textcomp}
\usepackage{xcolor}

\usepackage{bbding}

\def\BibTeX{{\rm B\kern-.05em{\sc i\kern-.025em b}\kern-.08em
    T\kern-.1667em\lower.7ex\hbox{E}\kern-.125emX}}
\begin{document}

% \title{Optimizing Irregular-Shaped Matrix-Matrix Multiplication on Multi-Core DSPs 
% }

\title{Optimizing Irregular-Shaped Matrix-Matrix Multiplication on Multi-Core DSPs
\thanks{The corresponding author is Qinglin Wang, and this work was supported by the National Natural Science Foundation of China under Grant No. 62002365.}
}

% \author{\IEEEauthorblockN{Shangfei Yin\IEEEauthorrefmark{1}, 
% 		Qinglin Wang\IEEEauthorrefmark{1}, 
% 		Ruochen Hao\IEEEauthorrefmark{1}, 
% 		Tianyang Zhou\IEEEauthorrefmark{1}, 
% 		Songzhu Mei\IEEEauthorrefmark{1} and 
% 		Jie Liu\IEEEauthorrefmark{1}}
% 	\IEEEauthorblockA{\IEEEauthorrefmark{1}Science and Technology on Parallel and Distributed Processing Laboratory,\\ National University of Defense Technology, Changsha 410073, China \\
% 		Email: yin\_shangfei@163.com, wangqinglin\_thu@163.com}
% }

\author{\IEEEauthorblockN{Shangfei Yin\IEEEauthorrefmark{1}\IEEEauthorrefmark{2}, Qinglin Wang\Envelope\IEEEauthorrefmark{1}\IEEEauthorrefmark{2}, Ruochen Hao\IEEEauthorrefmark{1}\IEEEauthorrefmark{2}, Tianyang Zhou\IEEEauthorrefmark{1}\IEEEauthorrefmark{2}, Songzhu Mei\IEEEauthorrefmark{1}\IEEEauthorrefmark{2}, Jie Liu\IEEEauthorrefmark{1}\IEEEauthorrefmark{2}}
	\IEEEauthorblockA{\IEEEauthorrefmark{1}Science and Technology on Parallel and Distributed Processing Laboratory\\
		National University of Defense Technology, Changsha 410073, China}
	\IEEEauthorblockA{\IEEEauthorrefmark{2}College of Computer, National University of Defense Technology, Changsha 410073, China}
	\IEEEauthorblockA{Email: yin\_shangfei@163.com, wangqinglin\_thu@163.com}}

\maketitle
%\IEEEpeerreviewmaketitle

\begin{abstract}
General Matrix Multiplication (GEMM) has a wide range of applications in scientific simulation and artificial intelligence. Although traditional libraries can achieve high performance on large regular-shaped GEMMs, they often behave not well on irregular-shaped GEMMs, which are often found in new algorithms and applications of high-performance computing (HPC). Due to energy efficiency constraints, low-power multi-core digital signal processors (DSPs) have become an alternative architecture in HPC systems. Targeting multi-core DSPs in FT-m7032, a prototype CPU-DSPs heterogeneous processor for HPC, an efficient implementation - ftIMM - for three types of irregular-shaped GEMMs is proposed. FtIMM supports automatic generation of assembly micro-kernels, two parallelization strategies, and auto-tuning of block sizes and parallelization strategies. The experiments show that ftIMM can get better performance than the traditional GEMM implementations on multi-core DSPs in FT-m7032, yielding on up to $7.2\times$ performance improvement, when performing on irregular-shaped GEMMs. And ftIMM on multi-core DSPs can also far outperform the open source library on multi-core CPUs in FT-m7032, delivering up to $3.1\times$ higher efficiency.
%General Matrix Multiplication (GEMM) has a wide range of applications in artificial intelligence, scientific simulation and other fields. Although traditional GEMM library can often achieve high performance in large-scale and regular-shaped matrix-matrix multiplication, it often cannot achieve high performance in irregular-shaped matrix-matrix multiplication, which is also widely utilized in convolutional neural networks(CNN), etc. Based on MATRIX2, we propose new implementation for irregular-shaped matrix-matrix multiplication. Since the realization of irregular-shaped matrix-matrix multiplication requires more kernels, we realize the automatic generation of kernels for different shapes. For different scales of irregular-shaped matrix-matrix multiplication, we propose parallelization strategies based on M dimension and K dimension respectively, which are optimized with methods such as double-buffer, reducing space overhead, and dynamic block sizes. We compare the performance of this method with traditional implementation on MATRIX2. The proposed method can achieve higher performance in irregular-shaped matrix-matrix multiplication.
\end{abstract}


\section{Introduction}
%%%%%%%%%%%%%%%%%%%%%%%%%%%%%%%%%%%%%GEMM Applications 
As a key routine of the Basic Linear Algebra Subroutines (BLAS) library, the general matrix-matrix multiplication (GEMM) has been extensively applied in various fields, such as scientific simulations, data analytic and deep learning. There are many well-known BLAS libraries \cite{openblas2012, GotoBlas, BLIS}, which provide highly optimized GEMMs for specific platforms. Unfortunately, the optimization has mainly focused on large regular-shaped matrices (where both dimensions of matrices are large and close to each other), which are the most common cases in the High Performance Linpack (HPL).

%%%%%%%%%%%%%%%%%%%%%%%%%%%%%%%%%%%%%Irregular-shaped Matrix
With the evolution of GEMM applications, the sizes and shapes of matrices involved in GEMM can often be notably changing when different algorithms and input data are carried out. In the field of scientific simulations, the application of Finite Element Method (FEM) in fluid dynamics generates many GEMMs working on small matrices \cite{heinecke2016libxsmm}. In the field of traditional machine learning, the implementation of classical K-means clustering algorithm calculates the distance between many samples and multiple centroids by the means of GEMM \cite{dhillon2004kernel}, where the sizes of matrices are determined by the number of samples, centroids, and dimensions for each sample. For most datasets \cite{drake2012accelerated, hamerly2010making}, the number of samples are much bigger than the numbers of centroids and dimensions, which are usually small, so that the input matrices in K-means are irregular-shaped matrices where one dimension is much larger than the other \cite{rivera2021tsm2x}. In the field of deep learning, the general implementation of convolutional layers often transforms convolution operations into equivalent GEMM by means of image-to-column method (known as im2col) \cite{jia2014caffe, ijcnnwang2019}. One dimension of the matrix generated by im2col is equal to the product of the number, height and width of output images, and the other is dependent on the product of the number of channels, height and width of filters in convolutional layers. For the first layers of most Convolutional Neural Networks (CNNs) \cite{vgg2014, resnet}, the former is much bigger than the latter. From the top to bottom layers in CNNs, the output images gradually become smaller, and the number of channels increases gradually, so the sizes and shapes of matrices in GEMM also varies greatly. 

%%%%%%%%%%%%%%%%%%%%%%%%%%%%%%%%%%%%%Prior Work
Recently, there have been many efforts focusing on the optimization of irregular-shaped GEMMs on GPUs and CPUs. Rivera and Chen et al. \cite{chen2019tsm2, rivera2020ism2, rivera2021tsm2x} proposed different algorithms for the optimization of irregular-shaped GEMMs with input matrices of various shapes on GPUs. Ernst \cite{ernst2021performance} et al. described the implementation of real and complex irregular-shaped GEMMs on GPUs. Tang \cite{tang2021efficient} et al. evaluated the performance of Tensor Core-based mixed-precision irregular-shaped GEMMs on GPUs. Yang et al. \cite{yang2021libshalom} first conducted the optimization for irregular-shaped GEMMs running on ARMv8 CPUs. Li et al. \cite{Li2021AutoTSMM} proposes an auto-tuning framework for irregular-shaped GEMMs on CPUs with different architectures.

%%%%%%%%%%%%%%%%%%%%%%%%%%%%%%%%%%%%%%%%%%DSP
Due to the constraints of energy efficiency and power consumption, low-power embedded architectures have been introduced into the heterogeneous computing domain in high-performance computing, such as digital signal processors (DSPs) \cite{igual2012unleashing, tiwari2018high, wang2021advancing}. Compared to CPUs and GPUs, DSPs often features Very Long Instruction Word (VLIW) or vector cores without out-of-order execution. What's more, cores in DSPs typically work on software managed on-chip memory, and integrates Direct Memory Access (DMA) engines for data transmission between on-chip software managed memory and off-chip main memory. As a result, existing optimizations for GEMMs on CPUs and GPUs are not directly fit for GEMMs on DSPs. While the efficient implementations of regular-shaped GEMMs targeting DSPs have been intensively studied in \cite{van2016blis, maDRAM2019, liu2018matrix}, there are few works focusing on the optimization of irregular-shaped GEMM on DSPs. 

%%%%%%%%%%%%%%%%%%%%%%%%%%%%%%%%%%%%%%%%%%Our work
FT-m7032 is one of the prototype CPU-DSPs heterogeneous processors explored by our university to accomplish general computing by means of DSPs. It integrates 16 ARMv8 CPU cores for running operating systems, and 4 multi-core general-computing DSP (GPDSP) clusters for providing maximum computing performance, detailed explained in Section \ref{FT-m7032-arc}. Targeting multi-core DSPs (also called GPDSP clusters) in FT-m7032, this paper presents an efficient implementation for irregular-shaped GEMMs. To the best of our knowledge, this is the first work which optimizes irregular-shaped GEMMs on multi-core DSPs. The main contributions of this paper are as follows. 

\begin{itemize}
    \item We analyze problems and challenges of optimizing irregular-shaped matrices on multi-core DSPs in detail. It is found that various micro-kernels and parallelization methods are indispensable for irregular-shaped GEMMs to achieve high performance.
    
    \item We propose a new implementation (ftIMM) for three types of irregular-shaped GEMMs on multi-core DSPs in FT-m7032. FtIMM integrates various micro-kernels by the automatic generation of micro-kernels, provides different parallelization strategies, and can automatically choose the optimal block sizes and the parallelization strategy for irregular-shaped input matrices by dynamic adjusting.
    
    \item The experiments on FT-m7032 show that ftIMM can provide various auto-generated micro-kernels which can achieve close to theoretical performance. For three types of irregular-shaped GEMMs, ftIMM can achieve a speedup of up to 7.2 times against the traditional efficient GEMM implementations \cite{maDRAM2019, liu2018matrix}. Further, ftIMM on multi-core DSPs can get more than $3.1\times$ improvement of efficiency compared with the OpenBLAS library \cite{openblas2012} running on multi-core CPUs on FT-m7032.
\end{itemize}

%%%%%%%%%%%%%%%%%%%%%%%%%%%%%%%%%%%%%%%%%%%%%%%%%%%%%%%%%%%%%%%%%%%

% \input{background}
% \input{background}

\section{FT-m7032 Heterogeneous Processor}
\label{FT-m7032-arc}
\begin{figure}[h]
	\centering
	\includegraphics[width=\linewidth]{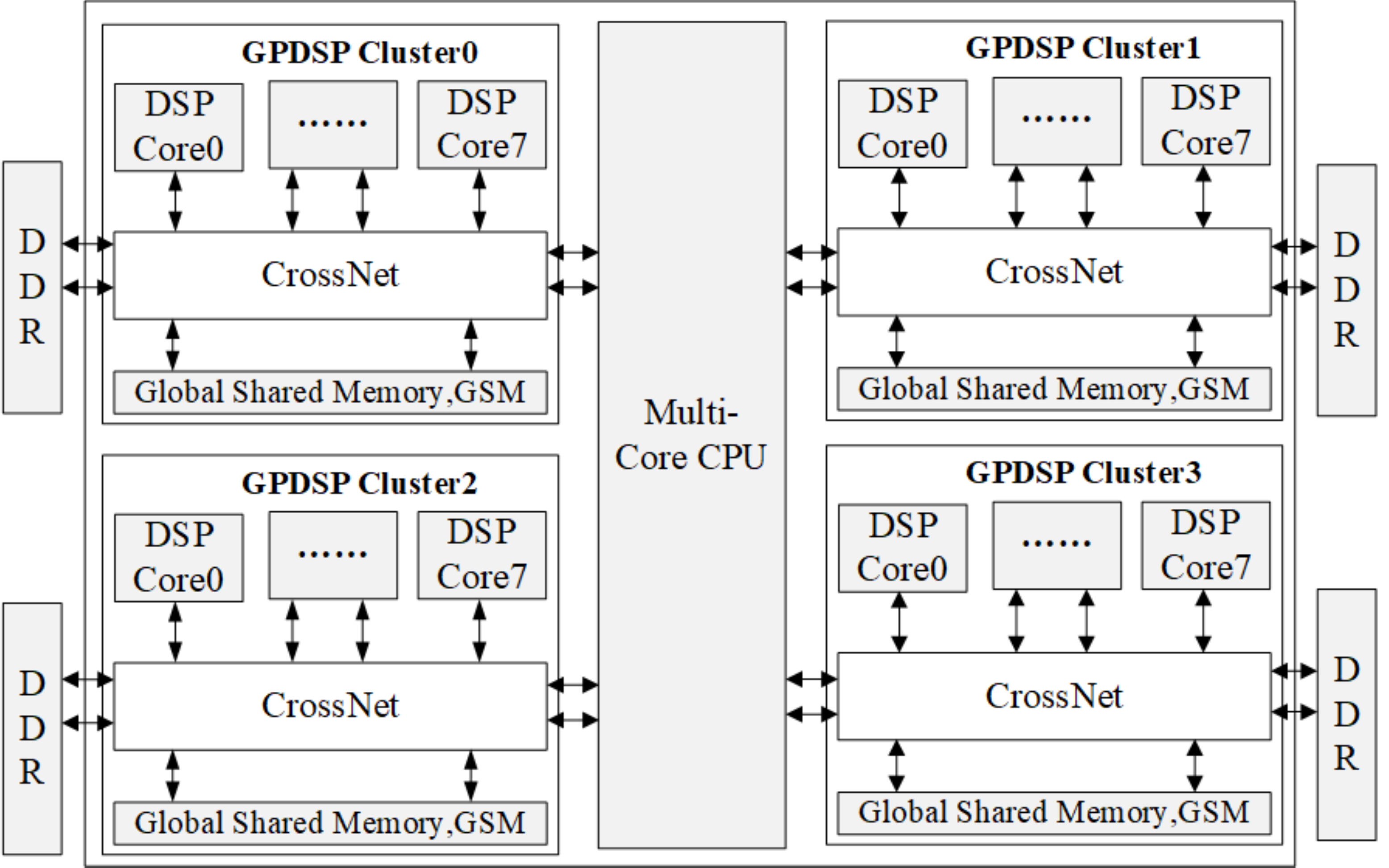}
	\caption{Architecture of FT-m7032}
	\label{fig:ft-m7032}
\end{figure}

%%%%%%%%%%%%%%%%%%%%%%%%%%%FT-m7032
The FT-m7032 heterogeneous processor consists of one 16-core ARMv8 CPU and four GPDSP clusters, as shown in Fig. \ref{fig:ft-m7032}. The multi-core CPU is a simplified version of Phytium FT-2000plus Processor\cite{HUANG2021102248, huang2021numa, wang2020OptimizingFFT}, and is chiefly responsible for process-level management and communication. The single-precision floating point peak performance of the multi-core CPU is 281.6 GFlops. Each GPDSP cluster includes eight DSP cores, which share 6 MB on-chip Global Shared Memory (GSM). All eight DSP cores and GSM in each cluster can communicate via a on-chip crossbar network, and the data coherency among them need to be maintained by software developers. The multi-core CPU and four GPDSP clusters share the same main memory space. The multi-core CPU can access the whole main space, but each GPDSP cluster can only access its own corresponding part with 42.6 GB/s hardware bandwidth. As the cache coherency among cores in the CPU is provided as in FT-2000plus, the cache data of CPU must be written back to the main memory before a function running on each GPDSP cluster starts, and be evicted after it finishes.
%FT-m7032 is high-performance floating-point multi-core vector processors developed by the National University of Defense Technology, which includes four GPDSP clusters and each cluster integrates eight vector processor cores, as shown in figure 1. The global shared memoty (GSM) and DDR are shared among multiple cores. The bandwidth between off-chip and on-chip is around 38.4GB/s. 

\begin{figure}[h]
	\centering
	\includegraphics[width=\linewidth]{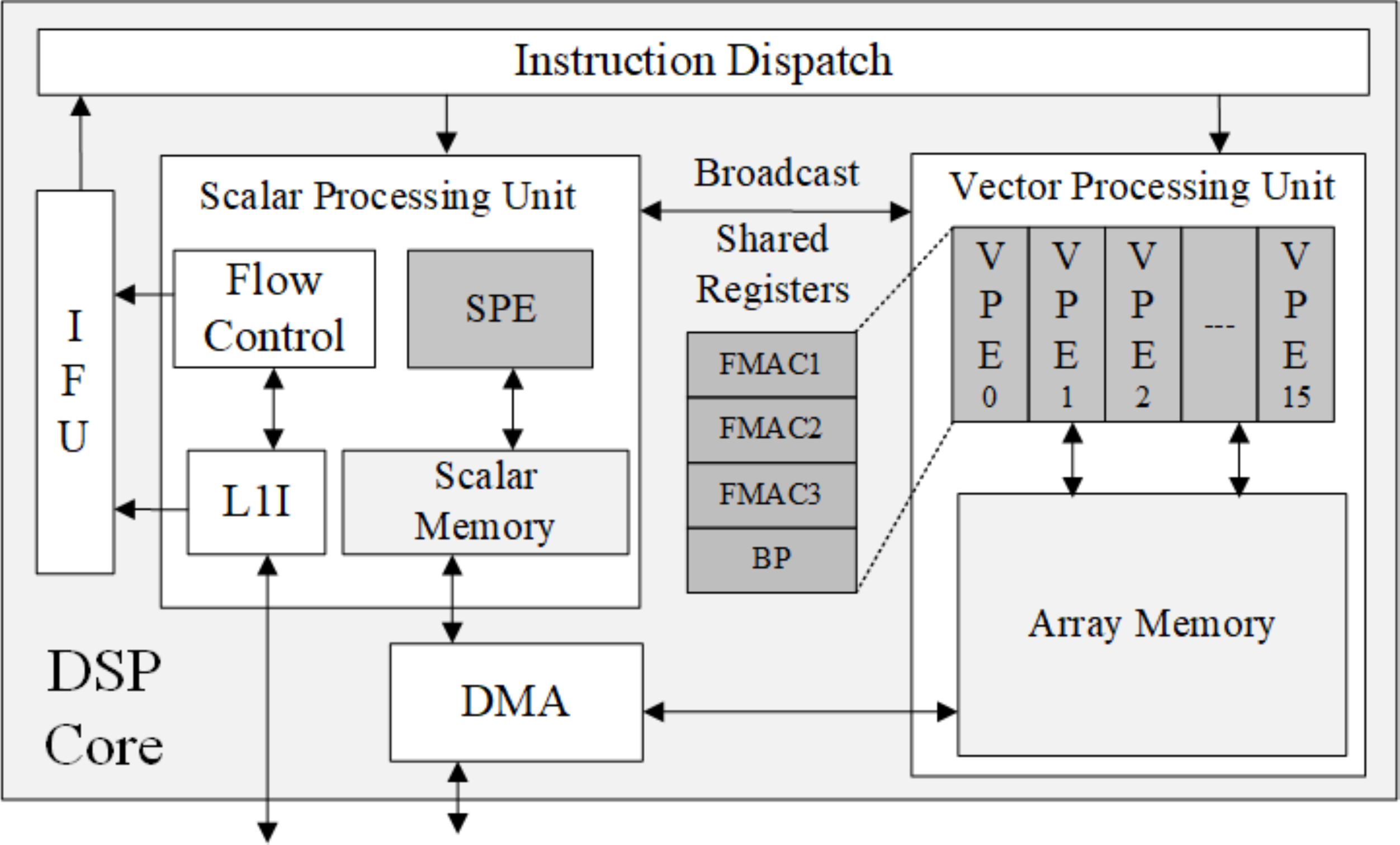}
	\caption{Micro-Architecture of DSP Core in FT-m7032}
	\label{fig:dsp-core}
\end{figure}

%%%%%%%%%%%%%%%%%%%%%%%%%%%DSP core
Each DSP core in each GPDSP cluster, based VLIW architecture, includes an instruction dispatch unit (IFU), a scalar processing unit (SPU), a vector processing unit (VPU) and a DMA engine, as shown in Fig. \ref{fig:dsp-core}. IFU is designed to launch up to 11 instructions per cycle, which incorporate 5 scalar instructions and 6 vector instructions. SPU is used for instruction flow control and the scalar computation, and mainly consists of Scalar Processing Element (SPE) and 64 KB Scalar Memory (SM), which match five scalar instructions. VPU provides the main computing performance for each DSP core, including 768 KB Array Memory (AM) and 16 vector processing elements (VPE) working in a SIMD manner. Each VPE has 64 64-bit registers and three fused multiply-add (FMAC) units, one of which can deal with two FP32 multiply-add computation per cycle. In other words, the SIMD width for FP32 data type is 32 and each DSP core can provide a peak performance of 345.6 GFlops when working at 1.8 Ghz. AM can deliver up to 512 bytes per cycle to registers by means of two load-store vector units. Between SPU and VPU, data can be transferred through the broadcast instruction and shared registers. The DMA engine is utilized to transfer data between different levels of memories (i.e. main memory, GSM, and SM/AM).

\section{Motivation}
\label{motivation}

\subsection{Irregular-shaped GEMMs}
\label{irs-gemm}
This paper mainly involves the optimization of single precision GEMMs with irregular-shaped matrices on a GPDSP cluster of FT-m7032, where at least one of $M$ and $K$ is sufficiently large and $N \leq 96$. In other word, for the three matrices ($A$, $B$ and $C$) in GEMMs ($C \; + = A \times B$), at least one matrix is tall-and-skinny (i.e., the height is significantly larger than the width). Specifically, there are three types of matrix-matrix multiplications: the multiplication between a tall-and-skinny matrix and a small matrix (i.e., $M \gg K \simeq N$), the multiplication between a skinny-and-tall matrix and a tall-and-skinny matrix (i.e., $K \gg M \simeq N$), and the multiplication between a large regular matrix and a tall-and-skinny matrix (i.e., $M \simeq K \gg N$).
%Based on the multi-core vector processors, the traditional GEMM implementation is mainly designed and optimized for large-scale and regular-shaped matrix-matrix multiplication. Although the implementation achieves high performance, it cannot perform well for irregular-shaped matrices. Therefore we focus on the optimization of the irregular-shaped GEMM, which refers to three types of GEMM in which $N \leq 96$ and M or K is large  and the traditional implementation cannot utilize multiple cores. Specifically, the three types of irregular-shaped GEMM are multiplication of a tall-and-skinny matrix and a small square matrix, multiplication of two tall-and-skinny matrices and multiplication of a large square matrix and a tall-and-skinny matrix. 

\subsection{Traditional GEMM Implementations}
The implementation of regular-shaped GEMMs based on \cite{maDRAM2019, liu2018matrix} is shown in Algorithm \ref{tgemm}, which is also based on the Goto GEMM algorithm \cite{GotoBlas} and labeled as TGEMM. Compared with the implementation on CPUs, there is no process of explicitly packing. In the following, matrices with subscripts $g$, $s$ and $a$ indicate that the matrices are located in GSM, SM and AM, respectively. The detailed process of TGEMM is described as follows.

%The implementation divides matrices into various sub-matrices based on the data locality of matrices in different levels of on-chip memory. The innermost computation within nested loops is the matrix multiplications carried on the sub-matrices, and the sub-matrices loaded into on-chip memory need to be reused as much as possible. 
 
The matrix $A$ is grouped into sub-matrices of size $m_g \times k_g$ by the outermost two loops (indexed by $i$ and $j$), which are first loaded into $A_g$ in GSM by DMA (Line \ref{tgemm:dmaA:gsm}). The first and third loops (indexed by $i$ and $t$) partition $C$ into sub-matrices of size $m_g \times n_a$, while the second and third loops (indexed by $j$ and $t$) split $B$ into sub-matrices of size $k_g \times n_a$. The sub-matrices of $B$ and $C$ are loaded into $B_a$ and $C_a$ by DMAs (Lines \ref{tgemm:dmaB:am} and \ref{tgemm:dmaC:am}), respectively. The fourth loop (indexed by $ii$) further divides the sub-matrices $A_g$ and $C_a$ into sub-matrices of sizes $m_s \times k_g$ and $m_s \times n_a$ , and then the $m_s \times k_g$ sub-matrices of $A_g$ are transferred into $A_s$ in SM by DMA (Line \ref{tgemm:dmaA:sm}). Thus the innermost computation (Line \ref{tgemm:inner}) (called as Micro-kernel) gets a $m_s \times n_a$ sub-matrix of $C_a$. After the fourth loop finishes, the sub-matrices $C_a$ of size $m_g \times n_a$ are stored back into $C$ in DDR by DMA. 

The loop order in Algorithm \ref{tgemm} is mainly determined by maximizing the reuse of sub-matrices on GSM, SM and AM. For the third and fourth loop (indexed by $t$ and $ii$), the sub-matrices $A_g$ and $B_a$ are repeatedly used in GSM and AM, respectively. The block sizes ($m_g$, $k_g$, $n_a$ and $m_s$) are chiefly resolved by maximizing the computation-to-memory ratio (CMR) of each level of on-chip memories under the capacity limit. In the case of single-precision GEMMs, the optimal parameters are that $m_g = 512$, $k_g = 512$, $m_s=6$, and $n_a = 96$. The implementation of micro-kernel is shown in Algorithm \ref{tgemm:micro-kernel}. The iterations in the third loop (indexed by $nn$) are mapped into three FMAC units in each VPE, and the second loop (indexed by $mm$) are unrolled to hide the latency of FMAC units. 

At the same time, the ping-pong method (also known as double-buffering method) is adopted to overlap computation and data transfer in Algorithm \ref{tgemm}. For example, the ping-pong method is performed on the iterations of the fourth loop (Lines \ref{tgemm:ii} - \ref{tgemm:inner}), so that the cost for the DMA loading (Line \ref{tgemm:dmaA:sm}) in $ii+1$th iteration can be overlapped with the overhead for the computation (Line \ref{tgemm:inner}) in $ii$th iteration. The parallelization on multi DSP cores are performed on the $N$ dimension (Line \ref{tgemm:t}) so that all cores shares the same sub-matrix $A_g$ in GSM.

\begin{algorithm}[htb!]
	\SetAlgoNoLine
	\SetAlgoNoEnd	
	\DontPrintSemicolon
	%\SetKwData{Left}{left}\SetKwData{This}{this}\SetKwData{Up}{up}
	\SetKwFunction{Union}{Union}
	\SetKwFunction{FindCompress}{FindCompress}
	\SetKwFunction{kernel}{KernelHrxWr}
	\SetKwProg{Fn}{Function}{:}{}
	\SetKwInOut{Input}{input}\SetKwInOut{Output}{output}
	\Input{$A[M][K]$, $B[K][N]$}
	\Output{$C[M][N]$}
	Set $m_g = 512$, $k_g = 512$, $n_a = 96$ and $m_s = 6$ \;
	\For{$i= 0 \colon m_{g} \colon M$\label{tgemm:i}} {
		\For(\tcp*[f]{Ping-pong}){$j = 0 \colon k_{g} \colon K $\label{tgemm:j}}  {
			DMA($A_{i,j}[m_g][k_{g}]$ $\to$ $A_g[m_g][k_{g}])$ \label{tgemm:dmaA:gsm}\;
			\ForPar(\tcp*[f]{Ping-pong}){$t = 0 \colon n_{a} \colon N $\label{tgemm:t}}  {
				DMA($B_{j,t}[k_g][n_{a}]$ $\to$ $B_{a}[k_g][n_{a}]$) \label{tgemm:dmaB:am}\;
				DMA($C_{i,t}[m_g][n_{a}]$ $\to$ $C_{a}[m_g][n_{a}]$) \label{tgemm:dmaC:am}\;
				\For(\tcp*[f]{Ping-pong}){$ii = 0 \colon m_{s} \colon m_{g} $\label{tgemm:ii}}  {
					DMA($A_{g,ii}[m_{s}][k_{g}]$ $\to$ $A_{s}[m_{s}][k_{g}]$) \label{tgemm:dmaA:sm}\;
					\tcp{Micro-kernel}
					$C_{a,ii}[m_{s}][n_{a}] += A_s[m_{s}][k_{g}] \times B_a[k_{g}][n_{a}]$ \label{tgemm:inner}
				}
				DMA($C_{a}[m_g][n_{a}]$ $\to$ $C_{i,t}[m_g][n_{a}]$)\label{tgemm:dmaC:ddr}\;
			}
		}
	}
	\caption{Traditional GEMM Implementation (TGEMM)}
	\label{tgemm}
\end{algorithm}

%%%%%%%%%%%%%%%%%%%%%%%MicroKernel
\begin{algorithm}[htb!]
	\SetAlgoNoLine
	\SetAlgoNoEnd	
	\DontPrintSemicolon
	\SetKwFunction{Union}{Union}
	\SetKwFunction{FindCompress}{FindCompress}
	\SetKwFunction{kernel}{KernelHrxWr}
	\SetKwProg{Fn}{Function}{:}{}
	\SetKwInOut{Input}{input}\SetKwInOut{Output}{output}
	\Input{$A_{s}[m_s][k_g]$, $B_a[k_g][n_a]$}
	\Output{$C_{a}[m_s][n_a]$}
	Set $V=32$, $v_n=(n_a + V - 1)/V$\;
	VPU loads $C_{a}[m_s][n_a]$ to vector registers $V_{c,0:m_s,0:v_n}$\;
	\For{$kk = 0 \colon 1 \colon k_g$} {
		\For(\tcp*[f]{hide the latency of FMAC units}){$mm = 0 \colon 1 \colon m_s$} {
			SPU loads $A_{s, mm, kk}$ to scalar register $R_{mm}$ \;
			SPU broadcasts $R_{mm}$ to vector register $V_{a,mm}$ \; 
			\For(\tcp*[f]{map to FMAC1-3}){$nn = 0 \colon 1 \colon v_n$} {
				VPU loads $B_{a, kk,nn \times V : (nn + 1) \times V}$ to vector register $V_{b,nn}$\;
				$V_{c,mm,nn} = V_{a,mm} \times V_{b,nn}$
			}
		}
	}
	VPU store $V_{c,0:m_s,0:v_n}$ back to $C_{a}[m_s][n_a]$
	\caption{The Implementation of Micro-Kernel in TGEMM}
	\label{tgemm:micro-kernel}
\end{algorithm}

\subsection{Analysis of Applying TGEMM to Irregular-shaped Matrices}
As shown in \cite{maDRAM2019, liu2018matrix}, TGEMM can get good performance on large regular-shaped matrices. However, there are several problems and challenges in the application of TGEMM to irregular-shaped matrices. 

First of all, only a micro-kernel in TGEMM is supported so that the additional overhead will be incurred when performing on tall-and-skinny and small matrices by implicitly padding. For example, when $N$ is smaller than $n_a$, TGEMM still stores $B_a$ in AM as a $k_g \times n_a$ matrix so that the memory space of AM are wasted. Then, the maximum value of $k_g$ will be limited, and the number of DMA operations for loading $B$ into AM will be greatly increased. It also maybe introduces additional computation overhead that the micro-kernel still performs $k_g \times n_a \times m_s$ operations. It is necessary to quickly support various micro-kernels by automatic generation, where $m_s$ and $n_a$ are arbitrary sizes under hardware constraints.

%First of all, for the irregular-shaped matrix-matrix multiplication, because sizes of several dimensions are too small or the block sizes of several dimensions are too small in edge cases, it is necessary to design optimized micro-kernels in which sizes of M and N are arbitrary and limited by the hardware. In the traditional implementation, padding is used to handle this case with only one micro-kernel, which is lack of special optimization for different sizes of M or N and brings additional computation and space overhead, resulting in lower performance. Besides, automatic generation of micro-kernels is also significant because it takes too much work to implement assembly code of different micro-kernels manually.

% In addition, for the irregular-shaped matrix-matrix multiplication, because the size of N dimension is too small or the block size of N dimension is too small in the boundary case, the data along the N dimension cannot fully fill the vector register during the process of vectorization. The existing methods handle this case with padding, which brings redundant space overhead.

Secondly, only the multi-core parallelization based on the $N$ dimension is adopted in TGEMM so that it can not efficiently utilize multiple DSP cores when performing on irregular-shaped matrices where $N$ is small. In order to take full advantage of all DSP cores in a GPDSP cluster, various parallelization strategies matched with the special shapes of matrices need to be introduced.

%Secondly, the traditional implementation adopts the parallelization strategy based on the N dimension, and achieves high performance for the computation of large-scale and regular-shaped matrix-matrix multiplication. However, for irregular-shaped matrix-matrix multiplication, for example, the size of N dimension is much smaller than the M dimension, the traditional method cannot fully utilize the computation ability of the multi-core vector processors. As to achieve higher efficiency and performance, according to the special shape of irregular-shaped matrix multiplication, the corresponding parallelization strategy should be designed to make full use of multiple cores.

Finally, how to choose the optimal block sizes and parallelization strategy is a challenge when different micro-kernels and parallelization strategies can be available. In other words, in order to improve the efficiency of irregular-shaped GEMMs, the block sizes ($m_g$, $k_g$, $n_a$ and $m_s$, etc.) and parallelization strategy should be dynamically adjusted according to the sizes and shapes of matrices, and the micro-kernels based on the dynamic block sizes ($m_s$ and $n_a$) are then auto-generated and called.  

%Lack of a mechanism to choose an appropriate block size and parallelization strategy

%in the traditional method, the block sizes are designed according to the hardware restrictions and the CMR. For the irregular-shaped matrix-matrix multiplication, due to the particularity of the matrices sizes, it is difficult to obtain highest computing performance. The block sizes should be dynamically designed according to the sizes of matrices to improve the efficiency of the algorithm and we choose the proper auto-generated micro-kernel based on the dynamic block sizes.

% \input{design}
% \input{design}

\section{ Design of ftIMM}
To efficiently handle three types of irregular-shaped matrix-matrix multiplications on multi-core DSPs described in Section \ref{irs-gemm}, this section proposes an efficient implementation: ftIMM. First, an automatic generation scheme of various assembly micro-kernels are designed to support matrices of various shapes in irregular-shaped GEMMs. Then we design the multi-core algorithm of irregular-shaped GEMMs with two parallelization strategies. Due to the special shapes of matrices in irregular-shaped GEMMs, we dynamically adjust the parallelization strategy and block sizes to make full use of multiple cores and reduce additional overhead.

%There are two important parts, kernel implementation and multi-core algorithm design, in the implementation of GEMM on the multi-core vector processors FT-m7032. The kernel is the basic building block of GEMM, which performs the computation of multiplication of sub-matrices on single core. In the traditional implementation of large-scale and regular-shaped GEMMs on FT-m7032, there is no need to design different kernels based on the shapes of GEMMs because the shapes of matrices are regular and large enough to amortize the additional overhead and the optimal kernel can be determined by the hardware restrictions, computation-to-memory ratio (CMR) and so on. But the traditional implementation can not achieve highest performance for irregular-shaped GEMMs, because the implementation is lack of special optimization of kernels for irregular-shaped GEMM and the additional overhead is too high. Besides, multi-core algorithm design, especially the parallelization strategy among multiply cores, is significant to make full use of the computation ability. But the traditional implementation of GEMM utilizes the only parallelization strategy along the N dimension, which cannot utilize enough cores for irregular-shaped GEMMs. Therefore, it is significant to design the special GEMM implementations for irregular-shaped matrices on the multi-core vector processors.

\subsection{Micro-Kernel Design and Generation}
%As mentioned above, for irregular-shaped GEMM, the sizes of some dimensions are too small or the block sizes of some dimensions are too small in the edge cases. In the traditional implementation, the sizes of micro-kernels are determined according to the restrictions of hardware and the CMR and only one micro-kernel is utilized. Using only one single kernel with fixed sizes results in additional computation and space overhead for irregular-shaped GEMM. Therefore, it is necessary to design micro-kernels with different sizes according to the sizes of the matrices. Since high-performance micro-kernels are often implemented by assembly manually, which brings a lot of work, we realize the automatic generation of micro-kernels with different sizes and different instruction sets on the multi-core vector processors.

%By analyzing the implementation of micro-kernels on the multi-core vector processors, we design three types of kernels to utilize the fused multiply-add units efficiently. The main micro-kernels are to compute $A \times B$ with $64 < N \leq 96$. The second and third types of micro-kernels are designed for edge cases that $32 < N \leq 64$ and $0 < N \leq 32$ respectively, in which the traditional implementation cannot utilize all fused multiply-add units units.

\subsubsection{Design Principals}
Our micro-kernel implementation aims to utilize three FMAC units in VPEs efficiently and minimize the cost of on-chip memory access ($SM$ and $AM$). We achieve the target by making full use of the instruction-level parallelism, scalar and vector registers, and the broadcasting bandwidth between SPU and VPU. In FT-m7032, SPU can broadcast at most two FP32 scalars to two vectors per cycle owing to the instruction conflicts. VPU can load up to 128 FP32 data into 4 vector registers per cycle and perform up to three fused multiply-add operations of vectors per cycle. As a result, the bandwidth between VPEs and AM and the computation ability can meet the needs for the irregular-shaped GEMMs with $N \leqslant 96$, while the broadcasting bandwidth between SPU and VPU is likely to become a bottleneck of performance. It is very important to improve the parallelism of broadcast operations in the design and auto-generation of micro-kernels.

Based on the principals above, the general implementation of micro-kernels in ftIMM is proposed, shown in Algorithm \ref{ftimm:micro-kernel}. Compared with micro-kernels in TGEMM, there are three main differences. The first one is that ftIMM adds two loops (indexed by $mu$ and $ku$) by loop tiling, which will be unrolled in the assembly implementation. The tiling sizes ($m_u$ and $k_u$) rely on how to make full use of three FMAC units in parallel and hide their latency, explained in Section \ref{generation}. The second one is that there is a reduction operation (Lines \ref{ftimm:mkernel:begin} - \ref{ftimm:mkernel:end}) in ftIMM when the tiling size $k_u$ is greater than 1. The last one is that the loading operations are carried out according to the size of $n_a$ by VPU so that the implicitly padding in TGEMM are no longer required and the space utilization efficiency of AM can be improved.

\begin{algorithm}[htb!]
	\SetAlgoNoLine
	\SetAlgoNoEnd	
	\DontPrintSemicolon
	\SetKwFunction{Union}{Union}
	\SetKwFunction{FindCompress}{FindCompress}
	\SetKwFunction{kernel}{KernelHrxWr}
	\SetKwProg{Fn}{Function}{:}{}
	\SetKwInOut{Input}{input}\SetKwInOut{Output}{output}
	\Input{$A_{s}[m_s][k_a]$, $B_a[k_a][n_a]$}
	\Output{$C_{a}[m_s][n_a]$}
	Set $V=32$, $v_n=(n_a + V - 1)/V$\;
	\For{$mm = 0 \colon m_u \colon m_s$} {
		VPU inits $V_{c,0:k_u,0:m_u,0:v_n}$ to zero\;
		\For{$kk = 0 \colon k_u \colon k_a$} {
			\For{$mu = 0 \colon 1 \colon m_u$}{
				\For{$ku = 0 \colon 1 \colon k_u$} {
					SPU loads $A_{s, mm + mu, kk+ku}$ to scalar register $R_{mu,ku}$ \;
					SPU broadcasts $R_{mu,ku}$ to vector register $V_{a,mu,ku}$ \; 
					\For{$nn = 0 \colon 1 \colon v_n$} {
						VPU loads $B_{a, kk + ku, nn \times V : (nn + 1) \times V}$ to vector register $V_{b, ku, nn}$\;
						$V_{c,ku,mu,nn} += V_{a,mm,ku} \times V_{b,ku,nn}$
					}
				}
			}
		}
		\For{$ku = 1 \colon 1 \colon k_u$\label{ftimm:mkernel:begin}} {
			$V_{c,0,0:m_u,0:v_n} += V_{c,ku,0:m_u,0:v_n}$\label{ftimm:mkernel:end}
		}
		VPU store $V_{c,0,0:m_u,0:v_n}$ back to $C_{a,mm}[m_u][n_a]$.
	}
	\caption{The Implementation of Micro-Kernel in ftIMM}
	\label{ftimm:micro-kernel}
\end{algorithm}

\subsubsection{Generation of Assembly Micro-kernels}
\label{generation}
Based on a given micro-kernel ($m_s$, $k_a$ and $n_a$), the key to generating an efficient assembly code is that how to map the computation in the micro-kernel on FMAC units in DSP cores and try to hide the executing period of these FMAC units. According to the size of $n_a$, micro-kernels in ftIMM are generated in the following two ways.

The first is the case where $64 < n_a \leq 96$. As the same in TGEMM, the parallelism in $n_a$ is mapped into three FMAC units in 16 VPEs. We use different schemes to hide latency of the FMAC units according to $m_s$. For the micro-kernel where $m_s$ is greater than the latency of FMAC instructions ($t_{fma}$), the tiling size $k_u$ is set to 1, and the $m_u$ can be as large as possible to hide the latency under the limitation of the total number of registers. For the micro-kernel where $m_s < t_{fma}$, $m_u$ is set to $m_s$ and $k_u$ is set to be greater than 1 for hiding the latency $t_{fma}$. The generated assembly pipeline for the micro-kernel with different $m_s$ (i.e., $m_s \geq t_{fma}$) based on the current instruction sets is shown in table \ref{table:case1}, in which $t_{VLDW}$ and $t_{SBR}$ represent the latency of the vector data loading instruction (i.e., VLDW and VLDDW) and the jump instruction (i.e., SBR) respectively. All three FMAC units are filled, and all the operations are pipelined, such as scalar data loading (i.e., SLDH and SLDW), scalar data extending (i.e., SFEXT32L and SBALE2H), broadcasting (i.e., SVBCAST and SVBCAST2), vector data loading and fused multiply-add instructions (i.e., VFMULAS32). 

\begin{table*}[!ht]
	\renewcommand\arraystretch{1.1}
	\scriptsize
	\caption{Assembly pipeline for the micro-kernel where $m_s \geq t_{fma}$ and $64 < n_a \leq 96$\label{table:case1}}
	\centering
	\begin{tabular}{lllllllll}
		\hline
		Cycle & 1           & 2             & ... & $m_u - t_{VLDW}+1$               & ... & $m_u - t_{SBR}+1$                   & ... & $m_u$ \\
		\hline
		Scalar Load\&Store1 & SLDH        & SLDH          & ... & SLDH                           & ... & SLDH                              & ... & SLDH \\ 
		Scalar FMAC1 & \textbar  SFEXTS32L    & \textbar SFEXTS32L      & ... & \textbar SFEXTS32L                       & ... & \textbar SFEXTS32L                          & ... & \textbar SFEXTS32L \\ 
		Scalar FMAC2 & \textbar SVBCAST     & \textbar SVBCAST       & ... & \textbar SVBCAST                        & ... & \textbar SVBCAST                           & ... & \textbar SVBCAST \\ 
		Vector Load\&Store1 & ~           & ~             & ~   & \textbar VLDDW                          & ~   & ~                                 & ~   & ~ \\
	    Vector Load\&Store2 & ~           & ~             & ~   & \textbar VLDW                           & ~   & ~                                 & ~   & ~ \\
		Vector FMAC1 & \textbar VFMULAS32  & \textbar VFMULAS32    & ... & \textbar VFMULAS32                     & ... & \textbar VFMULAS32                        & ... & \textbar VFMULAS32 \\
		Vector FMAC2 & \textbar VFMULAS32  & \textbar VFMULAS32    & ... & \textbar VFMULAS32                     & ... & \textbar VFMULAS32                        & ... & \textbar VFMULAS32 \\
		Vector FMAC3 & \textbar VFMULAS32  & \textbar VFMULAS32    & ... & \textbar VFMULAS32                     & ... & \textbar VFMULAS32                        & ... & \textbar VFMULAS32 \\
		Control unit & ~           & ~             & ~   & ~                              & ~   & \textbar SBR                               & ~   & ~ \\ 
		% ~           & ~             & ~                              & SUB                               & ... & ~ \\ 
		\hline
	\end{tabular}
\end{table*}

The second is the case where $0 < n_a \leq 64$. For the case, the parallelism in $n_a$ is not sufficient for three FMAC units in 16 VPEs. In order to improve the utilization of FMAC units per cycle, the tiling size $k_u$ should be set to be greater than 1. At the same time, $m_u$ is also set to be as large as possible for hiding the latency of FMAC units. The generated assembly pipelines for the micro-kernels with $32 < n_a \leq 64$ and $0 < n_a \leq 32$ are shown in table \ref{table:case2} and \ref{table:case3}, where $m_s = 6$. The SIEU represents the unit which only supports the fixed-point operation. 

\begin{table*}[!ht]
	\renewcommand\arraystretch{1.1}
	\scriptsize
	\caption{Assembly pipeline for the micro-kernel where $m_s = 6$ and $32 < n_a \leq 64$	\label{table:case2}}
	\centering
	\begin{tabular}{lllllllll}
		\hline
		Cycle & 1 & 2 & 3 & 4 & 5 & 6 & 7 & 8 \\
		\hline
		Scalar Load\&Store1 & SLDW    & SLDW     & SLDW     & ~        & SLDW    & SLDW     & SLDW     & ~\\ 
		Scalar FMAC1 & \textbar SFEXTS32L& \textbar SFEXTS32L & ~        & SFEXTS32L & \textbar SFEXTS32L& \textbar SFEXTS32L & ~       & SFEXTS32L \\ 
		Scalar FMAC2 & \textbar SVBCAST2 & \textbar SVBCAST2  & \textbar SVBCAST2  & ~        & \textbar SVBCAST2 & \textbar SVBCAST2  & \textbar SVBCAST2  & ~  \\ 
		SIEU & \textbar SBALE2H & \textbar SBALE2H  & ~        & \textbar SBALE2H  & \textbar SBALE2H & \textbar SBALE2H  & ~        & \textbar SBALE2H  \\ 
		Vector Load\&Store1 & ~       & ~        & ~        & ~        & ~       & ~        & ~        & \textbar VLDDW  \\
		Vector Load\&Store2 & ~       & ~        & ~        & ~        & ~       & ~        & ~        & \textbar VLDDW  \\
		Vector FMAC1 & \textbar VFMULAS32 & \textbar VFMULAS32 & \textbar VFMULAS32 & \textbar VFMULAS32 & \textbar VFMULAS32 & \textbar VFMULAS32 & \textbar VFMULAS32 & \textbar VFMULAS32 \\ 
		Vector FMAC2 & \textbar VFMULAS32 & \textbar VFMULAS32 & \textbar VFMULAS32 & \textbar VFMULAS32 & \textbar VFMULAS32 & \textbar VFMULAS32 & \textbar VFMULAS32 & \textbar VFMULAS32 \\ 
		Vector FMAC3 & \textbar VFMULAS32 & \textbar VFMULAS32 & \textbar VFMULAS32 & \textbar VFMULAS32 & \textbar VFMULAS32 & \textbar VFMULAS32 & \textbar VFMULAS32 & \textbar VFMULAS32 \\  
		Control unit & ~       & \textbar SBR      & ~        & ~        & ~       & ~        & ~        & ~  \\
		% ~       & ~        & SUB      & ~        & ~       & ~        & ~        & ~  \\
		\hline
	\end{tabular}
\end{table*}

\begin{table*}[!ht]
	\renewcommand\arraystretch{1.1}
	\scriptsize
	\caption{Assembly pipeline for the micro-kernel where $m_s = 6$ and $0 < n_a \leq 32$	\label{table:case3}}
	\centering
	\begin{tabular}{llllllll}
		\hline
		Cycle & 1           & 2             & 3          & 4          & 5           & 6          & 7 \\
		\hline
		Scalar Load\&Store1 & SLDW        & ~             & SLDW       & SLDW       & SLDW        & SLDW       & SLDW\\ 
		Scalar FMAC1 & \textbar SFEXTS32L    & ~             & \textbar SFEXTS32L   & \textbar SFEXTS32L   & \textbar SFEXTS32L    & \textbar SFEXTS32L   & \textbar SFEXTS32L\\ 
		Scalar FMAC2 & \textbar SVBCAST2     & SVBCAST2       & ~          & \textbar SVBCAST2    & \textbar SVBCAST2     & \textbar SVBCAST2    & \textbar SVBCAST2\\ 
		SIEU & \textbar SBALE2H     & ~             & \textbar SBALE2H    & \textbar SBALE2H    & \textbar SBALE2H     & \textbar SBALE2H    & \textbar SBALE2H\\ 
		Vector Load\&Store1 & &               &            &            &             & \textbar VLDW      &        \\
		Vector Load\&Store2 & &               &            &            &             & \textbar VLDW      &         \\
		Vector FMAC1 & \textbar VFMULAS32  & \textbar VFMULAS32    & \textbar VFMULAS32 & \textbar VFMULAS32 & \textbar VFMULAS32  & \textbar VFMULAS32 & ~ \\ 
		Vector FMAC2 & \textbar VFMULAS32  & \textbar VFMULAS32    & \textbar VFMULAS32 & \textbar VFMULAS32 & \textbar VFMULAS32  & \textbar VFMULAS32 & ~ \\  
		Control unit & \textbar SBR         &               &            &            &             &            &        \\
		% & SUB           &            &            &             &            &         \\
		\hline
	\end{tabular}
\end{table*}

%%%%% new
\subsubsection{Upper Bound Performance of Micro-Kernels}
The utilization of three FMAC units affects the performance of micro-kernels. Due to the limited number of units in SPU, we can only broadcast two FP32 scalar data to vector registers per cycle and therefore the size of $n_a$ affects the utilization of FMAC units. For the micro-kernels where $32 < n_a \leq 96$, because two elements of matrix $A_s$ can be broadcast to vector registers and three or two vectors can be loaded from matrix $B_a$ in one cycle, all three FMAC units can be utilized and the theoretical upper bound performance is close to $100\%$. For the micro-kernels where $0 < n_a \leq 32$, only one corresponding vector can be loaded from matrix $B_a$ per cycle, which matches a vector broadcasting from matrix $A_s$, and at most two FMAC units can be utilized. Therefore the corresponding upper bound performance is about $66.7\%$.

% For main micro-kernels in which $64 < N \leq 96$, we can broadcast one element of matrix A to vector register in one cycle and multiply it by a row of matrix B which can utilize all three fused multiply-add units. Therefore the upper bound performance of main micro-kernels is $100\%$. 

\subsection{Multi-core Design}
\subsubsection{Parallelization on $M$ dimension}
The multi-core parallelization strategy based on $M$ dimension is designed and shown in the Algorithm \ref{multiCoreParaM}. The three-level DMA-based ping-pong scheme is designed to compute the matrix-matrix multiplication. Firstly, block in the $N$ and $K$ dimension, and perform the multiplication of the $M \times k_{g}$ sub-matrix of $A$ and the $k_{g} \times n_{g}$ sub-matrix of $B$ with the ping-pong in the loop (indexed by $j$). Block further in the $M$, $n_g$ and $k_g$ dimensions, perform ping-pong in the $k_g$ dimension based on DMA and perform the multiplication of the $m_{a} \times k_{a}$ sub-matrix of $A$ and the $k_{a} \times n_{a}$ sub-matrix of $B$. The multiplication of the $m_{s} \times k_{a}$ sub-matrix of $A$ and the $k_{a} \times n_{a}$ sub-matrix of $B$ is performed based on the ping-pong in the $m_a$ dimension and we can choose the corresponding auto-generated micro-kernel. In this strategy, we parallelize the loop of $M$ dimension to utilize multiple cores for irregular-shaped GEMM. Besides, GSM is used to cache the sub-matrix of $B$ which is shared among multiple cores. During the computation, each core loads its own private data of $A$ and $C$ from DDR and the shared sub-matrix of $B_g$ from GSM, which can fully realize data reuse and reduce the overhead of memory access. 

\begin{algorithm}[htb!]
	\SetAlgoNoLine
	\SetAlgoNoEnd	
	\DontPrintSemicolon
	\SetKwFunction{Union}{Union}
	\SetKwFunction{FindCompress}{FindCompress}
	\SetKwFunction{kernel}{KernelHrxWr}
	\SetKwProg{Fn}{Function}{:}{}
	\SetKwInOut{Input}{input}\SetKwInOut{Output}{output}
	\Input{$A[M][K]$, $B[K][N]$}
	\Output{$C[M][N]$}
	\For{$i = 0 \colon n_{g} \colon N$} {		
		\For(\tcp*[f]{Ping-pong}){$j = 0 \colon k_{g} \colon K$\label{ftIMM:i}}  {
			DMA($B_{j,i}[k_g][n_{g}]$ $\to$ $B_g[k_g][n_g]$)\label{ftIMM:dmaB:gsm}\; 			
			\ForPar{$t = 0 \colon m_{a} \colon M$\label{ftIMM:t}}{
				\For{$ii = 0 \colon n_{a} \colon n_g$}{
					DMA($C_{t,i + ii}[m_a][n_a]$ $\to$ $C_a[m_a][n_a]$)\label{ftIMM:dmaC:am}\;
					\For(\tcp*[f]{Ping-pong}){$jj = 0 \colon k_{a} \colon k_g$\label{ftIMM:jj}}{
						DMA($B_{g, jj, ii}[k_a][n_{a}]$ $\to$ $B_a[k_a][n_a]$)\label{ftIMM:damB:am}\;						
						\For(\tcp*[f]{Ping-pong}){$tt = 0 \colon m_{s} \colon m_a$\label{ftIMM:tt}}{
							DMA($A_{t + tt,j + jj}[m_s][k_a]$ $\to$ $A_s[m_s][k_a]$)\label{ftIMM:dmaA:sm}\;							
							$C_a[m_s][n_a] += A_s[m_s][k_a] \times B_a[k_a][n_a]$\label{ftIMM:inner}
						}
					}
					DMA($C_a[m_a][n_a]$ $\to$ $C_{t,i + ii}[m_a][n_a]$)\;
				}
			}
		}
	}
	\caption{Irregular-Shaped GEMM Implementation with Parallelization on M Dimension\label{multiCoreParaM}}
\end{algorithm}

\subsubsection{Parallelization on $K$ dimension}
The multi-core parallelization strategy based on $K$ dimension is shown in the Algorithm \ref{multiCoreParaK}. Similar to the $M$-dimension-based parallelization strategy, this strategy uses a two-level DMA-based ping-pong scheme to implement matrix-matrix multiplication. First, in the $K$ dimension, based on the ping-pong method, perform the multiplication of the $m_{a} \times k_{a}$ sub-matrix of $A$ and the $k_{a} \times n_{a}$ sub-matrix of $B$. Based on ping-pong in the $M$ dimension, perform the multiplication of the $m_{s} \times k_{a}$ sub-matrix of $A$ and the $k_{a} \times n_{a}$ sub-matrix of $B$. We parallelize the loop of $K$ dimension and use GSM to cache the sub-matrix of $C$, and perform the reduction between multiple DSP cores based on GSM, which can effectively realize data reuse and reduce the memory access overhead caused by reduction. Besides, because the parallelization strategy based on $K$ dimension brings additional overhead of reduction, this strategy is suitable for the multiplication of irregular-shaped GEMM with both small sizes of $M$ and $N$ dimensions, and therefore we do not utilize ping-pong in the outermost loop.

\begin{algorithm}[htb!]
	\SetAlgoNoLine
	\SetAlgoNoEnd	
	\DontPrintSemicolon
	\SetKwData{Left}{left}\SetKwData{This}{this}\SetKwData{Up}{up}
	\SetKwFunction{Union}{Union}
	\SetKwFunction{FindCompress}{FindCompress}
	\SetKwFunction{kernel}{KernelHrxWr}
	\SetKwProg{Fn}{Function}{:}{}
	\SetKwInOut{Input}{input}\SetKwInOut{Output}{output}
	\Input{Matrix $A[M][K]$, $B[K][N]$}
	\Output{Matrix $C[M][N]$}
	\For{$i= 0 \colon m_{g} \colon M$\label{ftIMM2:i}} {
		\For{$j = 0 \colon n_{g} \colon N$\label{ftIMM2:j}}  {
			DMA($C_{i,j}[m_g][n_g]$ $\to$ $C_g[m_g][n_g]$)\;		
			\For{$ii = 0 \colon m_{a} \colon m_g$}{
				\For{$jj = 0 \colon n_{a} \colon n_g$}{
				    Init sub-matrix $C_a[m_a][n_a]$ to zero\;
				% 	DMA($C_{g, i + ii, j + jj}[m_a][n_a]$ $\to$ $C_a[m_a][n_a]$)\label{ftIMM2:dmaC:am}\;
					\ForPar(\tcp*[f]{Ping-pong}){$t = 0 \colon k_{a} \colon K$\label{ftIMM2:t}}{
						DMA($B_{t, j + jj}[k_a][n_a]$ $\to$ $B_a[k_a][n_a]$)\label{ftIMM2:dmaB:am}\;	
						\For(\tcp*[f]{Ping-pong}){$u = 0 \colon m_{s} \colon m_{a}$\label{ftIMM2:u}}{
							DMA($A_{i + ii + u,t}[m_s][k_a]$ $\to$ $A_s[m_s][k_a]$)\label{ftIMM2:dmaA:sm}\;				
							$C_a[m_a][n_a] += A_s[m_a][k_a] \times B_a[k_a][n_a]$\label{ftIMM2:inner}
						}
					}		
					Reduce sub-matrix $C_a[m_a][n_a]$ among multiple cores based on GSM, and store the reduction result back into $C_{i+ii,j+jj}[m_a][n_a]$\;
				}
			}
		}
	}
	\caption{Irregular-Shaped GEMM Implementation with Parallelization on K Dimension}
	\label{multiCoreParaK}
\end{algorithm}

\subsection{Dynamic Adjusting}
In the traditional implementation of GEMMs on multi-core DSPs, the block sizes are designed according to the hardware constraints and CMR, and both the block sizes and the parallelization strategy are fixed whatever the shapes of matrices are. For large-scale regular-shaped GEMMs, the method in TGEMM works well. But for irregular-shaped GEMMs, due to special shapes, fixed parallelization strategy with fixed block sizes cannot make full use of the computation ability, and too large block sizes bring additional computation and space overhead, so the parallelization strategy and block sizes need to be adjusted dynamically according to the shape of matrices.

First, we design the initial block sizes according to the hardware constraints and CMR. In the parallelization strategy based on the $M$ dimension, the following two types of data transfer between multi-level memories are considered, and the corresponding CMR is calculated.

(1) The $k_{g} \times n_{g}$ sub-matrix of matrix $B$ is cached in GSM, the $m_{a} \times k_{g}$ sub-matrix of matrix $A$ is transferred between DDR and SM, and the $m_{a} \times n_{g}$ sub-matrix of matrix $C$ is transferred between DDR and AM. The CMR is shown in Eq. \ref{eq.1}.

(2) The $k_{a} \times n_{a}$ sub-matrix of matrix $B$ and $m_{a} \times n_{a}$ sub-matrix of matrix $C$ have been loaded to the AM, and the $m_{a} \times k_{a}$ sub-matrix of matrix $A$ is transferred between DDR and SM. The CMR is shown in Eq. \ref{eq.2}.

\begin{equation} \label{eq.1}    
	f_1=\frac{2 \times m_{a} \times k_{g} \times n_{g} \times num_{core}}{num_{core} \times m_{a} \times (k_{g} + 2 \times n_{g})+k_{g} \times n_{g}}
\end{equation}

\begin{equation} \label{eq.2}	    
	f_2=\frac{2\times m_{a}\times k_{a} \times n_{a} \times num_{core}}{num_{core} \times m_{a}\times (k_{a} + 2 \times n_{a}) + k_{a}\times n_{a}}
\end{equation}

Besides, when we design initial block sizes and adjust block sizes dynamically, we ensure the block size $k_{g}$ to be larger as much as possible because larger $k_{g}$ improves the reuse of the sub-matrix $C_a$ in AM which is loaded in line \ref{ftIMM:dmaC:am} of Algorithm \ref{multiCoreParaM}. Considering the space limit of GSM, SM and AM, to maximize the CMR, we can obtain the initial optimal block sizes for single-precision floating-point GEMM. The block sizes are that $k_{g} = 5888$, $n_{g} = 96$, $m_{a} = 320$, $n_{a} = 96$, $k_{a} = 864$ and $m_{s} = 8$.

Similar to the parallelization based on the $M$ dimension, the initial block sizes of the parallelization based on the $K$ dimension can be obtained through CMR and hardware constraints. The CMR are shown in Eq. \ref{eq.3} and \ref{eq.4}. The result is that $m_{g} = 1024$, $n_{g} = 512$, $m_{a} = 1024$, $n_{a} = 96$, $k_{a} = 512$ and $m_{s} = 14$.

\begin{equation} \label{eq.3}	    
	f_3=\frac{2 \times m_{g} \times k_{a} \times n_{g} \times num_{core}}{num_{core} \times k_{a} \times (m_{g} + n_{g}) + 2 \times m_{g} \times n_{g}}
\end{equation}

\begin{equation} \label{eq.4}		    
	f_4=\frac{2\times m_{a}\times k_{a} \times n_{a} \times num_{core}}{num_{core} \times k_{a}\times (m_{a} + n_{a}) + 2 \times m_{a}\times n_{a}}
\end{equation}

% Based on the auto-generation of micro-kernels of different shapes, we can generate the needed kernels of any sizes.
% Therefore 
After determining the initial block sizes, at runtime, ftIMM dynamically adjusts the parallelization strategy and block sizes based on the matrices sizes. For cases of multiplication between a tall-and-skinny matrix and a small matrix or multiplication between a large regular matrix and a tall-and-skinny matrix, in which $N \leq n_a$ and $M$ is large sufficiently, the parallelization strategy based on $M$ dimension is adopted. For the case of multiplication between a skinny-and-tall matrix and a tall-and-skinny matrix, in which $N \leq n_a$, $M$ is small enough and $K$ is large sufficiently, ftIMM chooses the parallelization strategy based on $K$ dimension which can utilize all cores in this case. Then the block sizes are adjusted based on the sizes of matrices and initial block sizes. For the case where the sizes are significantly smaller than the block sizes, ftIMM adjusts the block sizes to adapt to sizes of matrices, and increases the block size in the dimension selected for parallelization, to reduce additional computation and space overhead, and minimize memory access latency. Because micro-kernels with too small $m_s$ give lower performance than others, especially when $m_s \leq 6$, ftIMM ensures that $m_{s} \geq 6$ if $M$ is large enough during the adjusting of block sizes. 

% \bibliographystyle{IEEEtran}
% \bibliography{Ref-20220513}

% \input{experiment}
% \input{experiment}

\begin{figure*}[htb]
    \centering
    \subfigure[$N = 96, K = 512$]{
        \begin{minipage}[t]{0.3\linewidth}
            \centering
            \includegraphics[width=1\textwidth]{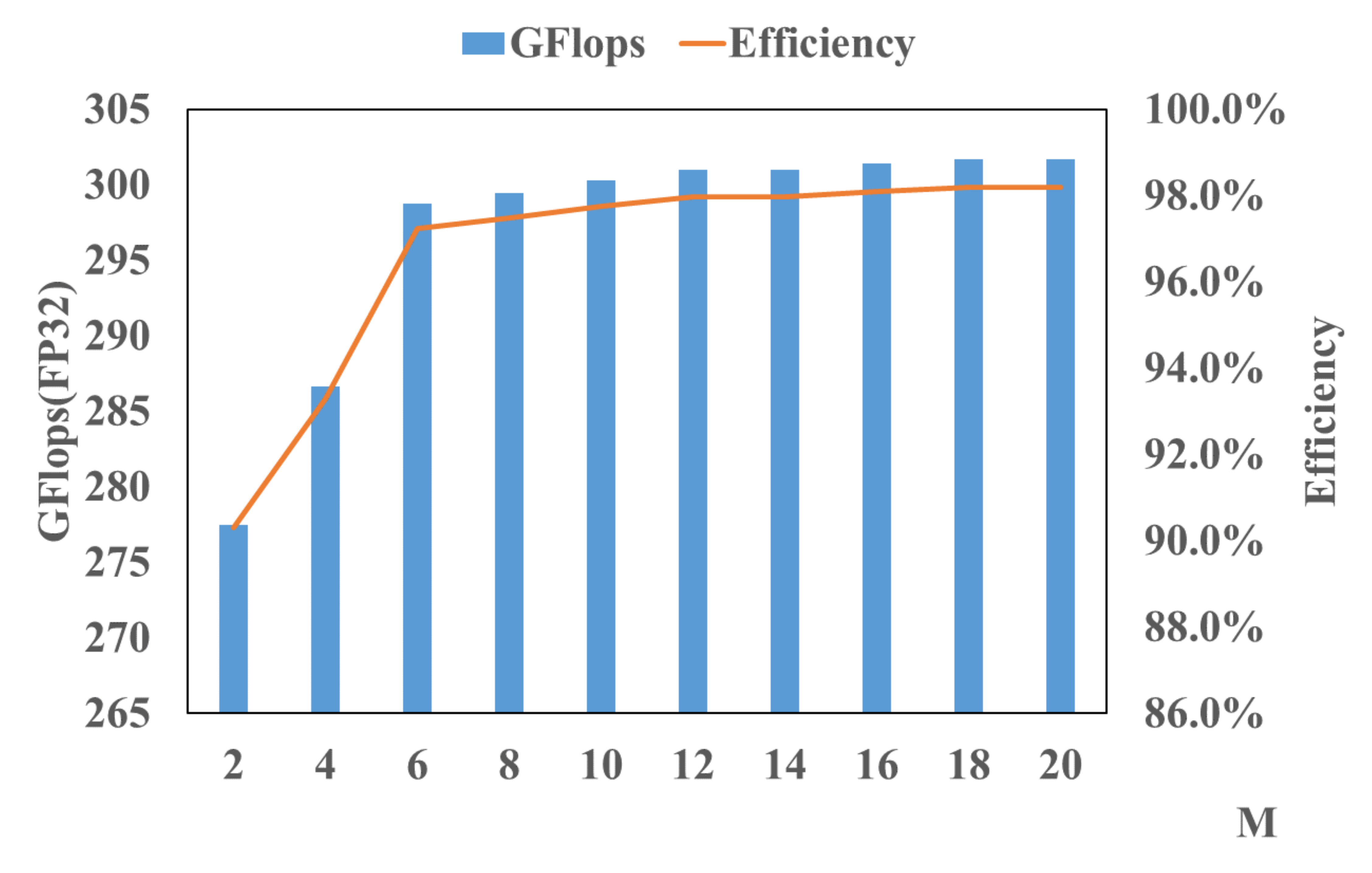}
            \end{minipage}%
            \label{fig:kernel-sf1}
    }
    \subfigure[$N = 64, K = 512$]{
        \begin{minipage}[t]{0.3\linewidth}
            \centering
            \includegraphics[width=1\textwidth]{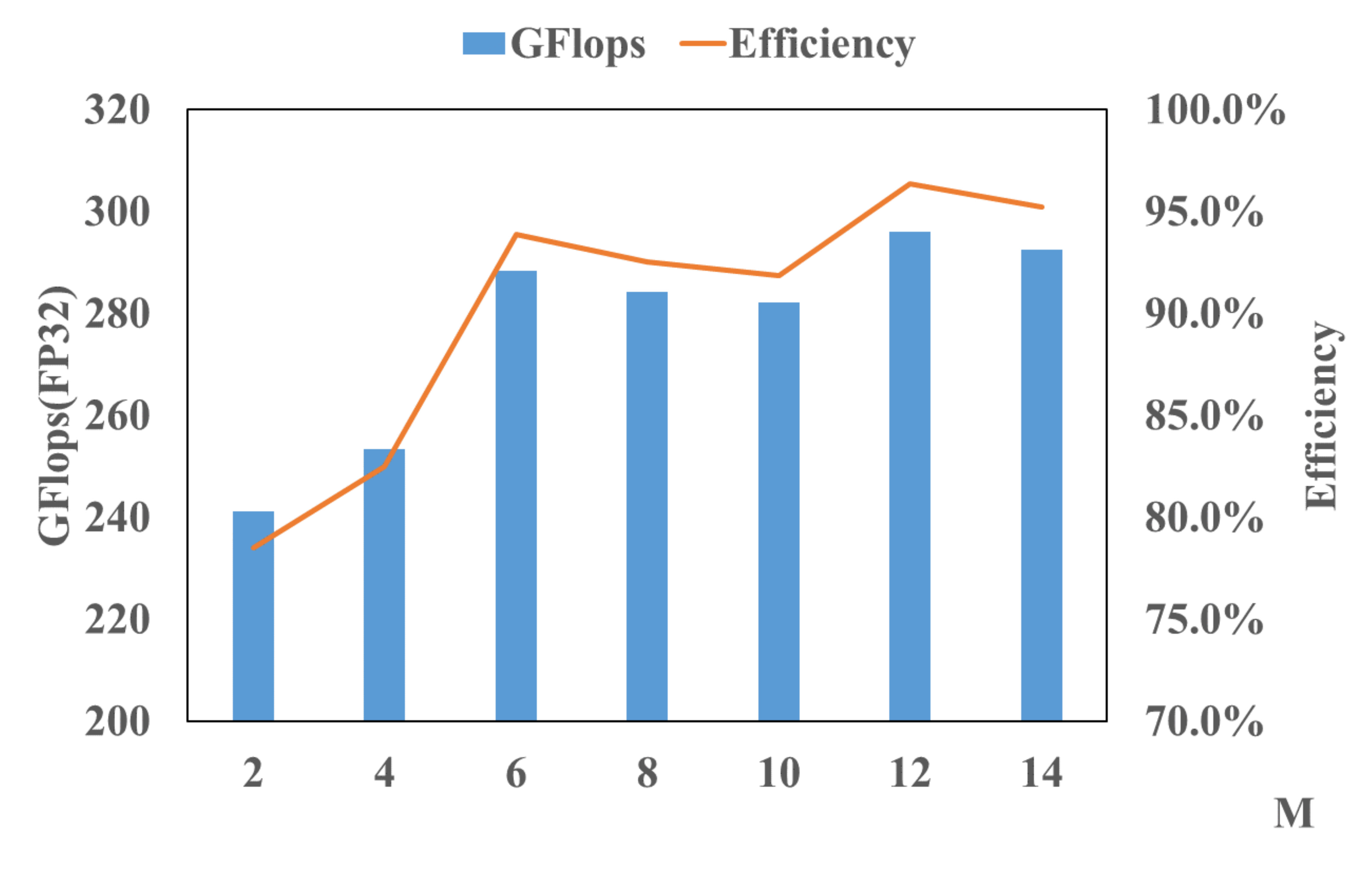}
            \end{minipage}%
            \label{fig:kernel-sf2}
    }
    \subfigure[$N = 32, K = 512$]{
        \begin{minipage}[t]{0.3\linewidth}
            \centering
            \includegraphics[width=1\textwidth]{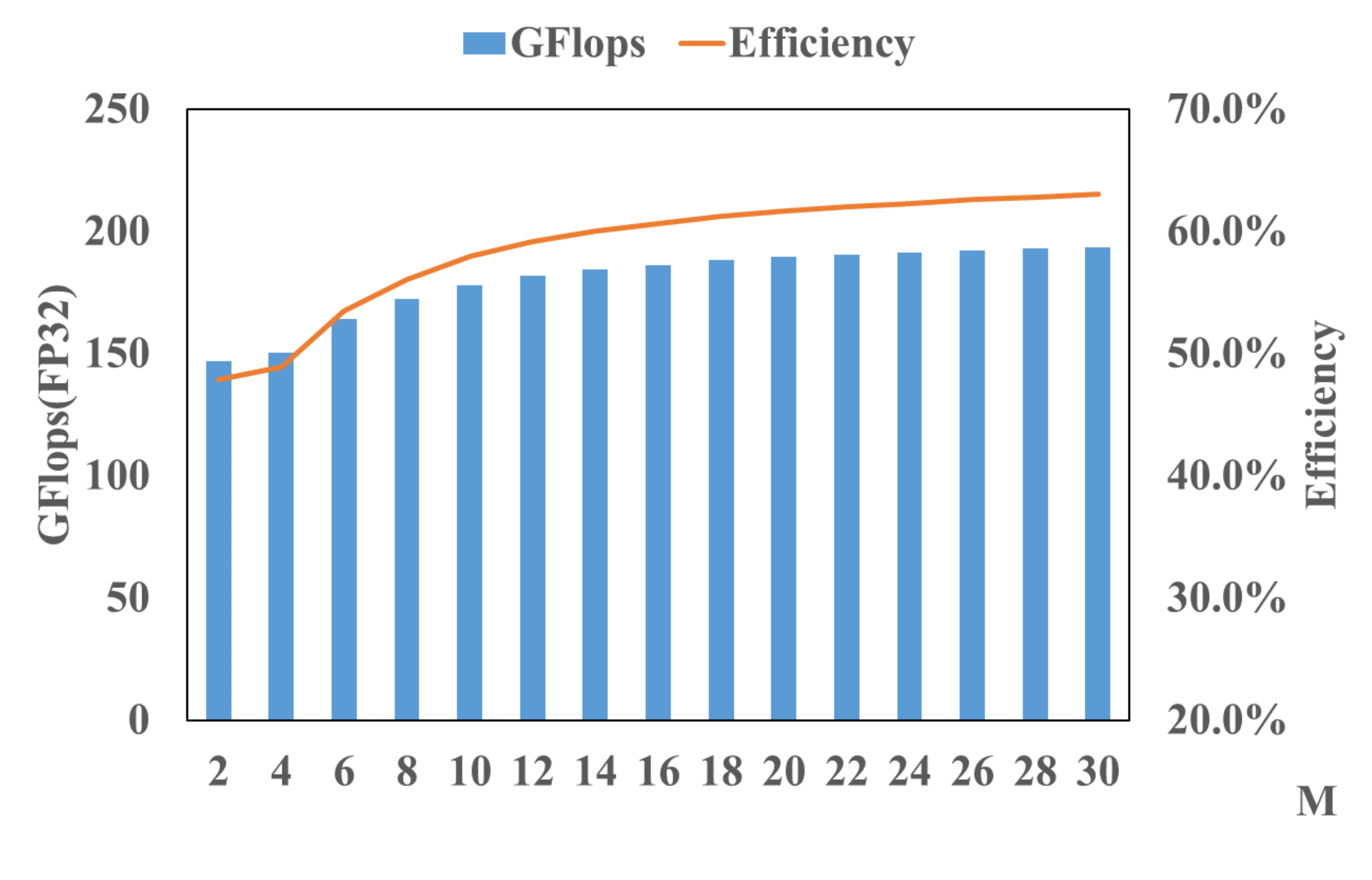}
            \end{minipage}%
            \label{fig:kernel-sf3}
    }
    
    \subfigure[$N = 96, K = 32$]{
        \begin{minipage}[t]{0.3\linewidth}
            \centering
            \includegraphics[width=1\textwidth]{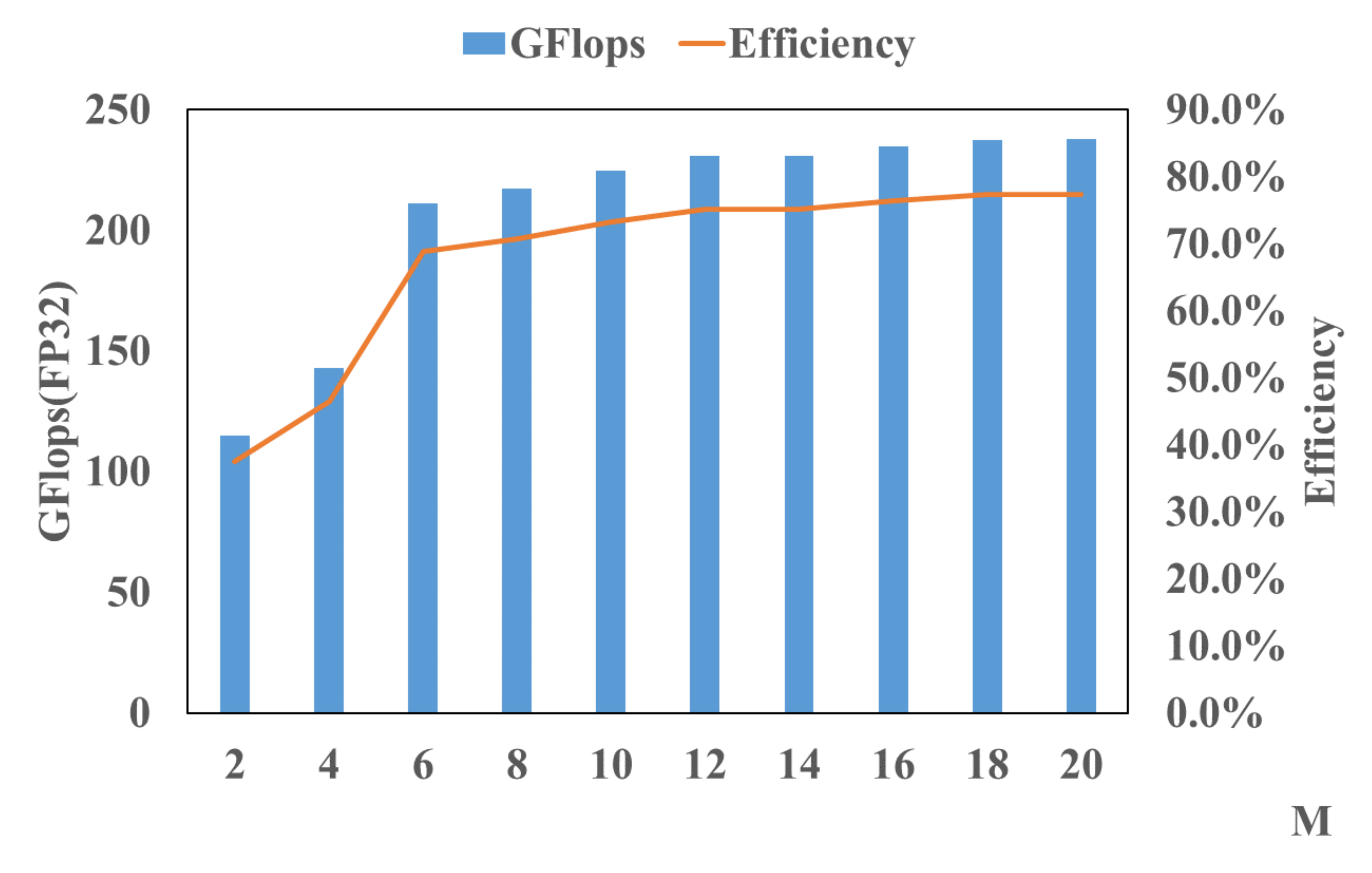}
            \end{minipage}%
            \label{fig:kernel-sf4}
    }%
    \subfigure[$N = 64, K = 32$]{
        \begin{minipage}[t]{0.3\linewidth}
            \centering
            \includegraphics[width=1\textwidth]{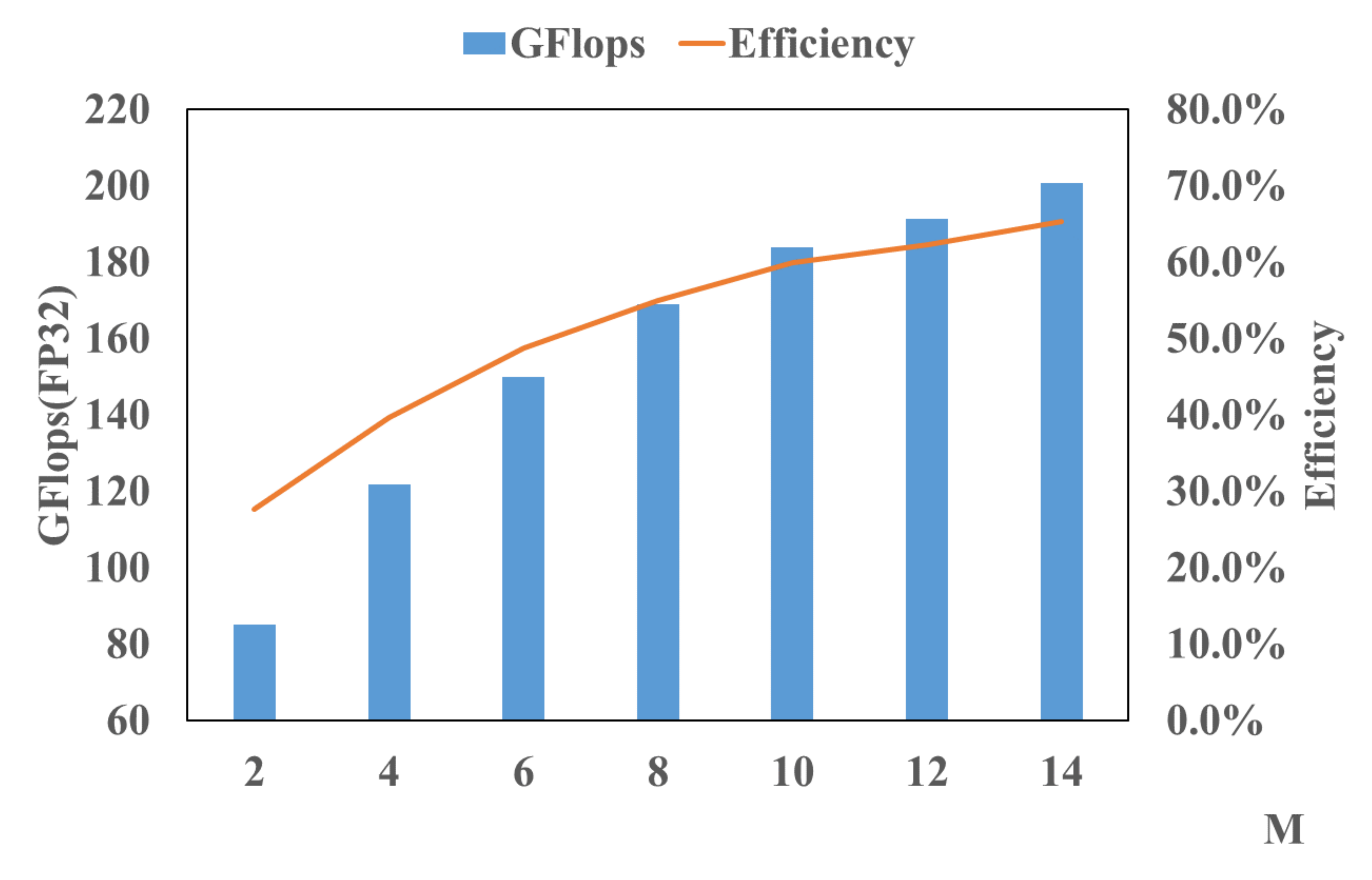}
            \end{minipage}%
            \label{fig:kernel-sf5}
    }
    \subfigure[$N = 32, K = 32$]{
        \begin{minipage}[t]{0.3\linewidth}
            \centering
            \includegraphics[width=1\textwidth]{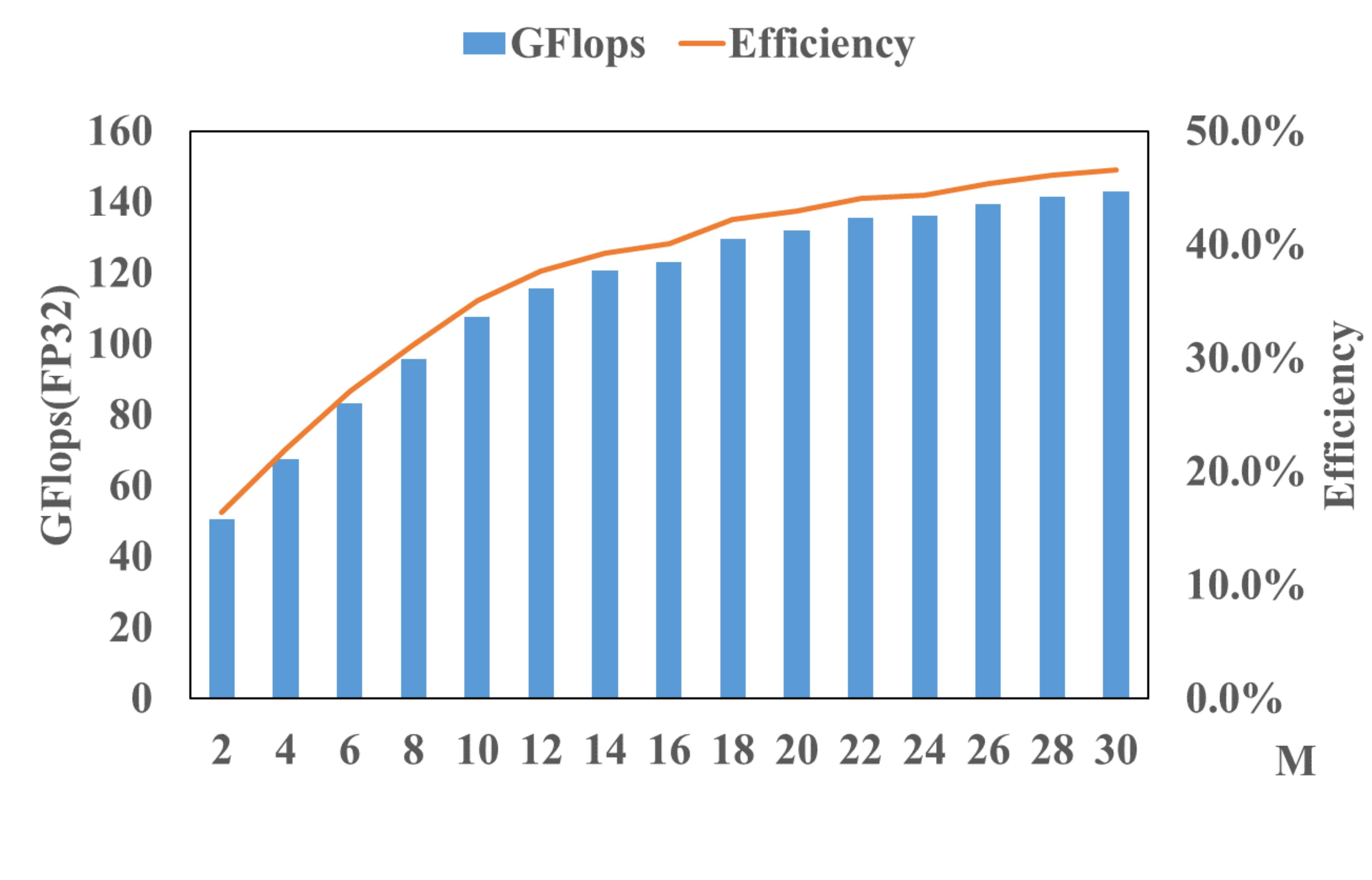}
            \end{minipage}%
            \label{fig:kernel-sf6}
    }
    \caption{Micro-Kernel Performance}
    \label{fig:kernel}
\end{figure*}

\section{Experiment}
\subsection{Micro-Kernel Performance}
We evaluate performance of auto-generated micro-kernels with single-precision workloads on two cases which are commonly used in the three types of irregular-shaped GEMMs. The first case is that K is large sufficiently (i.e., $K = 512$) and micro-kernels of this case are used in the multiplication between a skinny-and-tall matrix and a tall-and-skinny matrix and the multiplication between a large regular matrix and a tall-and-skinny matrix. The second case is that K is small enough (i.e., $K = 32$) which is used in the multiplication between a tall-and-skinny matrix and a small matrix. Because the size of M is limited by the hardware constraints and the specific implementation of micro-kernels, the size of M is different among the experiment of the two types of micro-kernels.

We first evaluate the performance of auto-generated micro-kernels where the size of K is fixed at 512. The results are shown in Fig. \ref{fig:kernel-sf1}, \subref{fig:kernel-sf2} and \subref{fig:kernel-sf3} in which $N = 96, 64, 32$ respectively. In these cases, the auto-generated micro-kernels achieve great performance. The highest is close to the theoretical upper bound, and the corresponding efficiency is up to $98.2\%$, $96.4\%$ and $63.0\%$ respectively. In Fig. \ref{fig:kernel-sf2}, micro-kernels with $M = 8, 10$ give lower performance than the micro-kernel with $M = 6$ and the micro-kernel with $M = 14$ gives lower performance than the micro-kernel with $M = 12$. This is mainly because in the implementations for micro-kernels in which $32 < N \leq 64$, there are not enough parallel FMAC operations to fill up the pipelines of FMAC units if $M\mod3 \neq 0$ and therefore micro-kernels with $M\mod3 \neq 0$ give lower performance.

We also evaluate the performance of the micro-kernels in which the size of K is small and the results are shown in Fig. \ref{fig:kernel-sf4}, \subref{fig:kernel-sf5} and \subref{fig:kernel-sf6}. The efficiency of cases in which $N = 96$ and $K = 32$ is up to $77.4\%$ which is high enough though K is too small. Besides, the efficiency for cases in which $N = 64$ and $N = 32$ can reach up to $65.4\%$ and $46.6\%$ which are close to or higher than the upper bound efficiency of micro-kernels in TGEMM respectively.

\subsection{Single-Core Performance}
\begin{figure*}[htb]
    \centering
    \subfigure[$M=2^{16}$]{
        \begin{minipage}[t]{0.3\linewidth}
            \centering
            \includegraphics[width =1\textwidth]{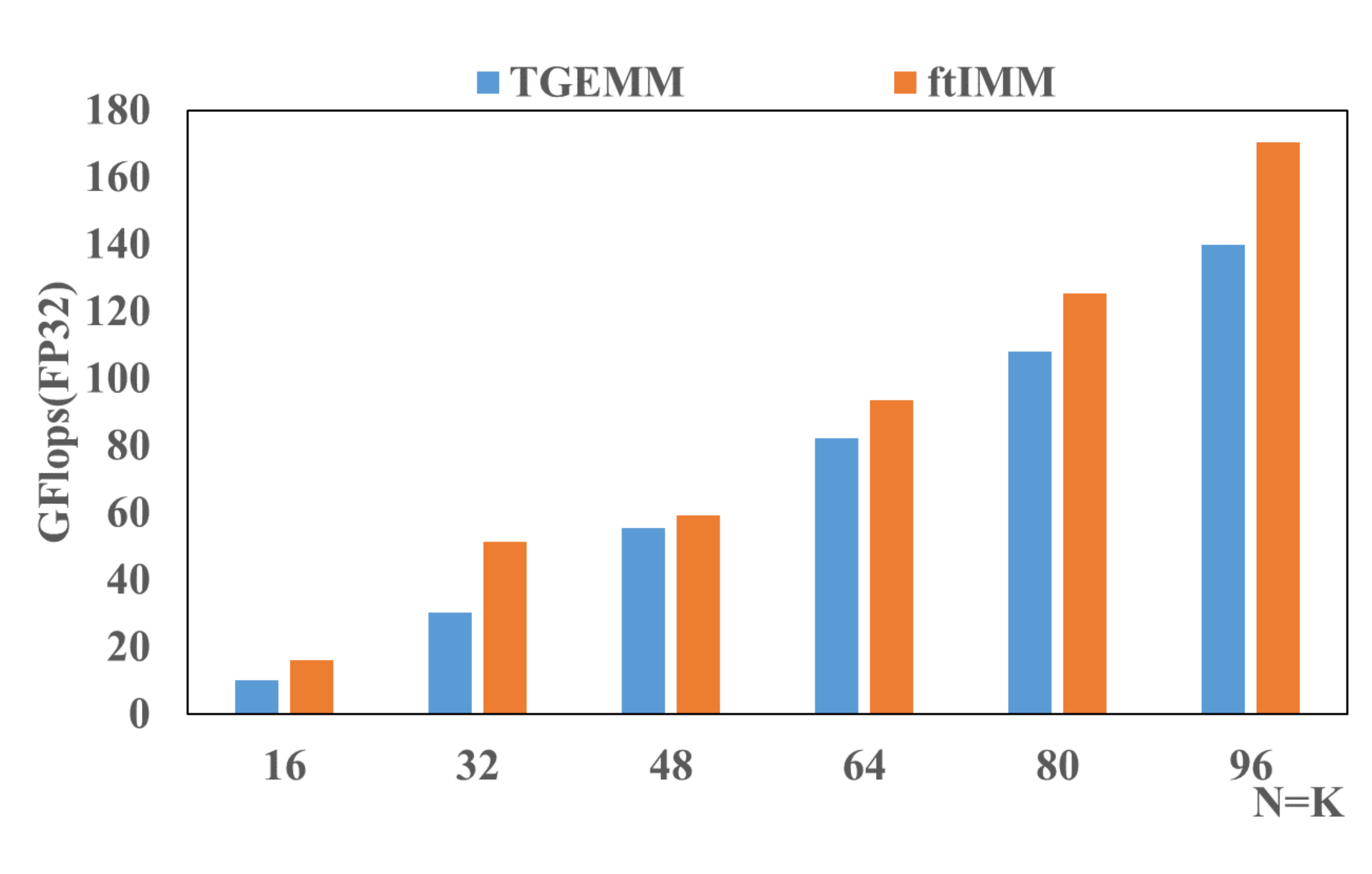}
            \end{minipage}%
            \label{fig:SingleCorePerformance-sf1}
    }
    \subfigure[$K=2^{16}$]{
        \begin{minipage}[t]{0.3\linewidth}
            \centering
            \includegraphics[width =1\textwidth]{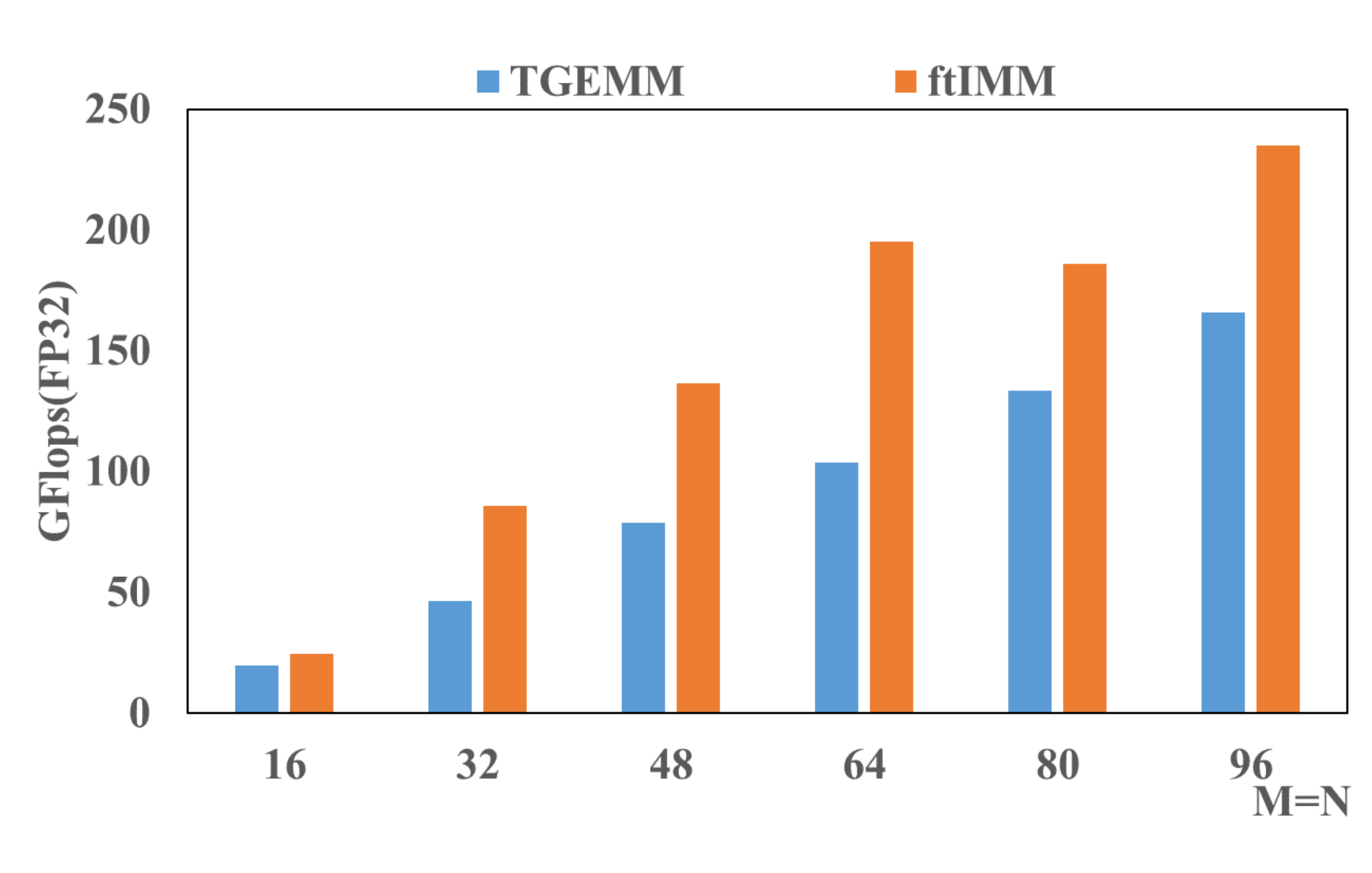}
            \end{minipage}%
            \label{fig:SingleCorePerformance-sf2}
    }
    \subfigure[$M=K=20480$]{
        \begin{minipage}[t]{0.3\linewidth}
            \centering
            \includegraphics[width =1\textwidth]{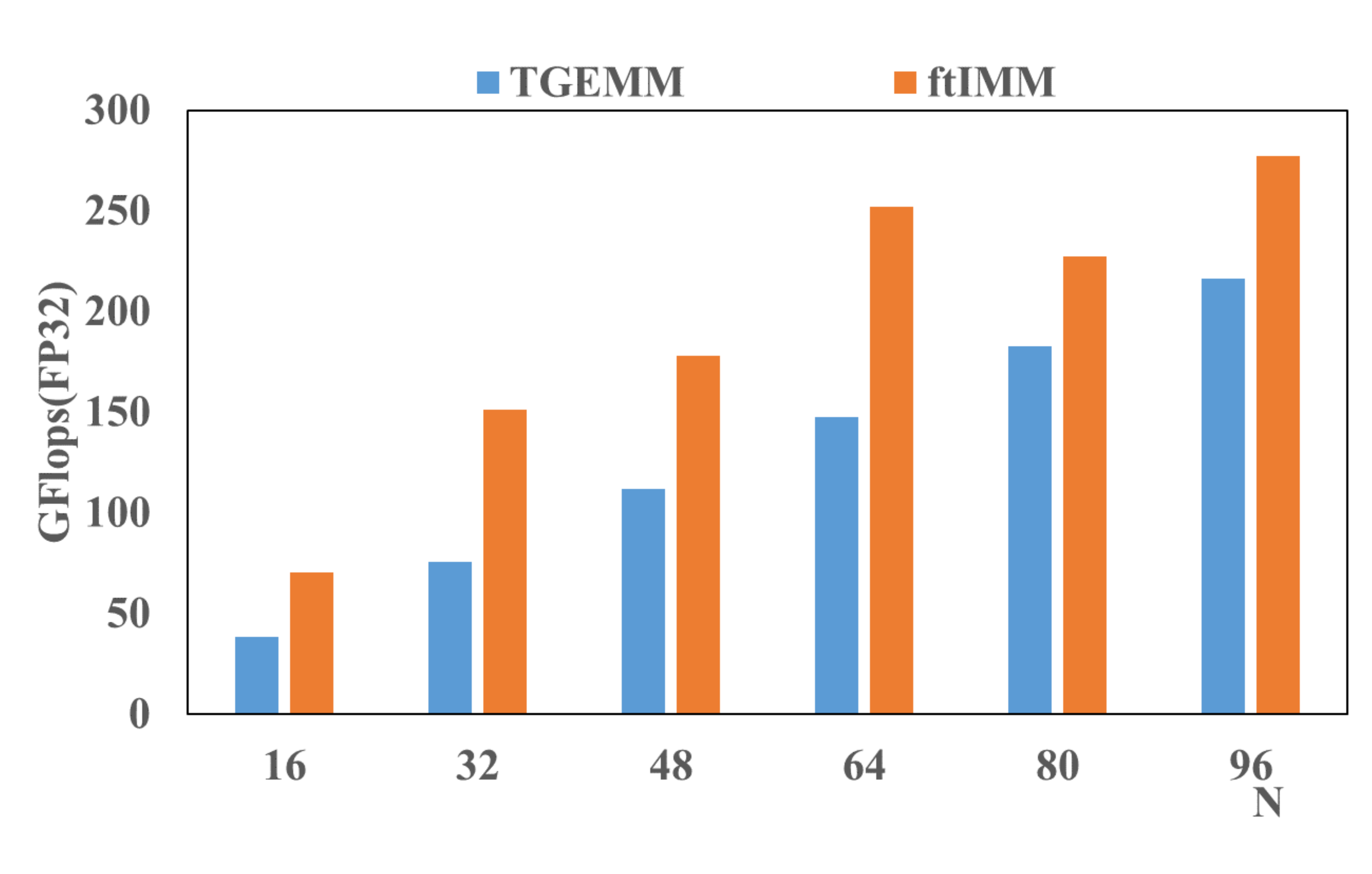}
            \end{minipage}%
            \label{fig:SingleCorePerformance-sf3}
    }
    \caption{Performance on Single DSP Core}
    \label{fig:SingleCorePerformance}
\end{figure*}

The performance of ftIMM is first evaluated by performing three types of irregular-shaped GEMMs on single DSP core. The results are shown in Fig. \ref{fig:SingleCorePerformance}. In all cases, ftIMM outperforms the traditional implementation TGEMM. For example, when $M \times N \times K = 20480 \times 32 \times 20480$, ftIMM gives $2.0\times$ higher performance than TGEMM. The improvement is especially obvious for the case in which the size of N dimension is much lower because of the optimized micro-kernels and dynamic adjusting. Specially, in \subref{fig:SingleCorePerformance-sf2} and \subref{fig:SingleCorePerformance-sf3}, ftIMM gives lower performance in the case where $N = 80$ than in the case where $N = 64$. This is because in the case where $N = 80$, block sizes, $m_a$, $k_g$ and so on, are smaller than those in the case where $N = 64$ and the performance is affected by block sizes. 

%  For the case in which the size of N dimension is large enough, the improvement is mainly because that larger block sizes of some dimensions reduce the number of DMA operations.

% The performance benefit is largely due to optimized micro-kernels and dynamic block sizes.
%%%%%

% The first case is $C += A \times B$ where A is a tall-and-skinny matrix and B is a small square matrix. The second case is $C += A^T \times B$ where A and B are both tall-and-skinny matrices. The third case is $C += A \times B$ where A is a large square matrix and B is a tall-and-skinny matrix. These types of irregular-shaped matrices are commonly used in convolution neural networks, K-means algorithm and so on. 

\begin{figure*}[htb]
    \centering
    \subfigure[$M=2^{16}$]{
        \begin{minipage}[t]{0.3\linewidth}
            \centering
            \includegraphics[width =1\textwidth]{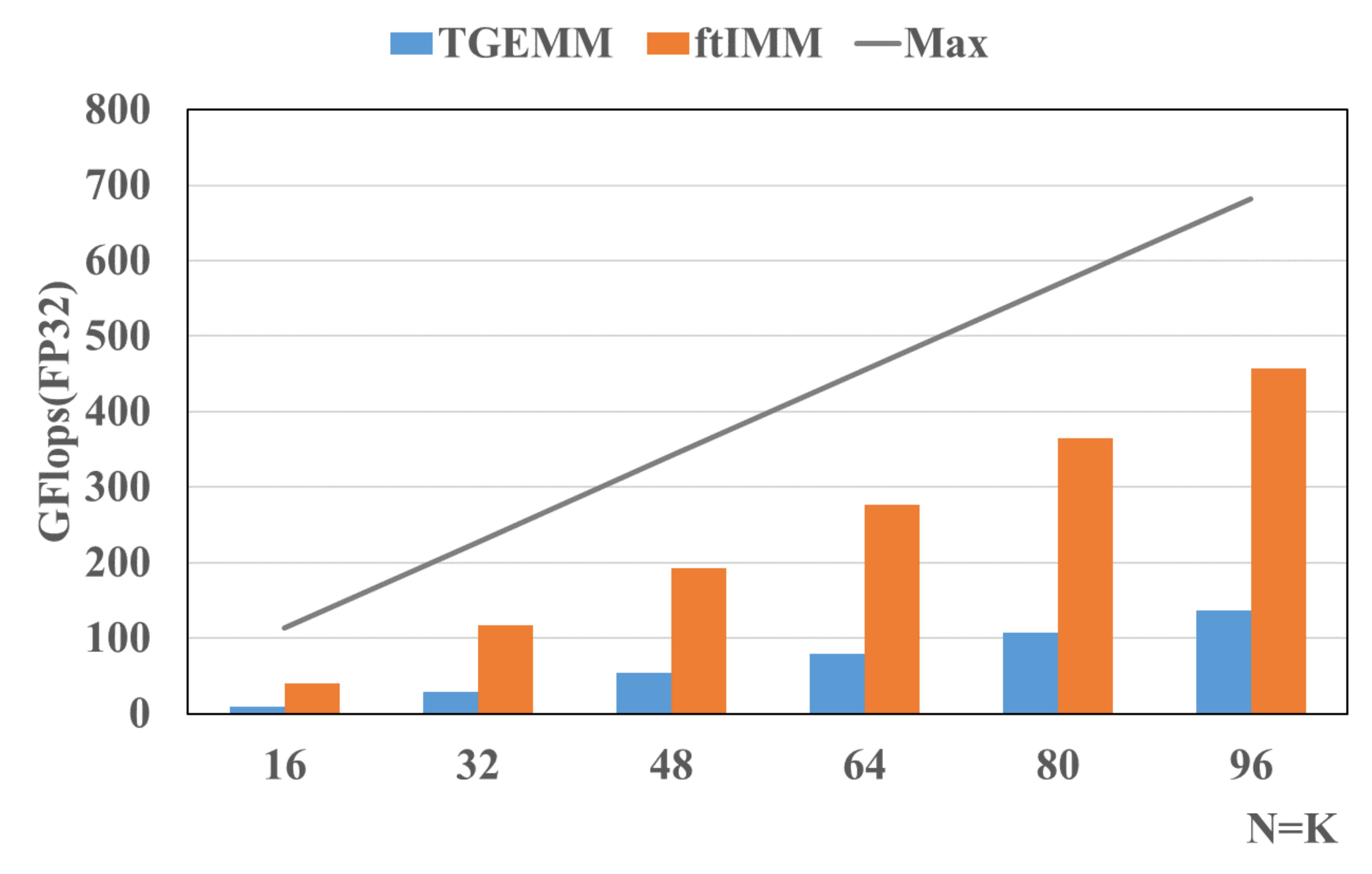}
            \end{minipage}%
            \label{fig:MultiCorePerformance-sf1}
    }
    \subfigure[$K=2^{16}$]{
        \begin{minipage}[t]{0.3\linewidth}
            \centering
            \includegraphics[width =1\textwidth]{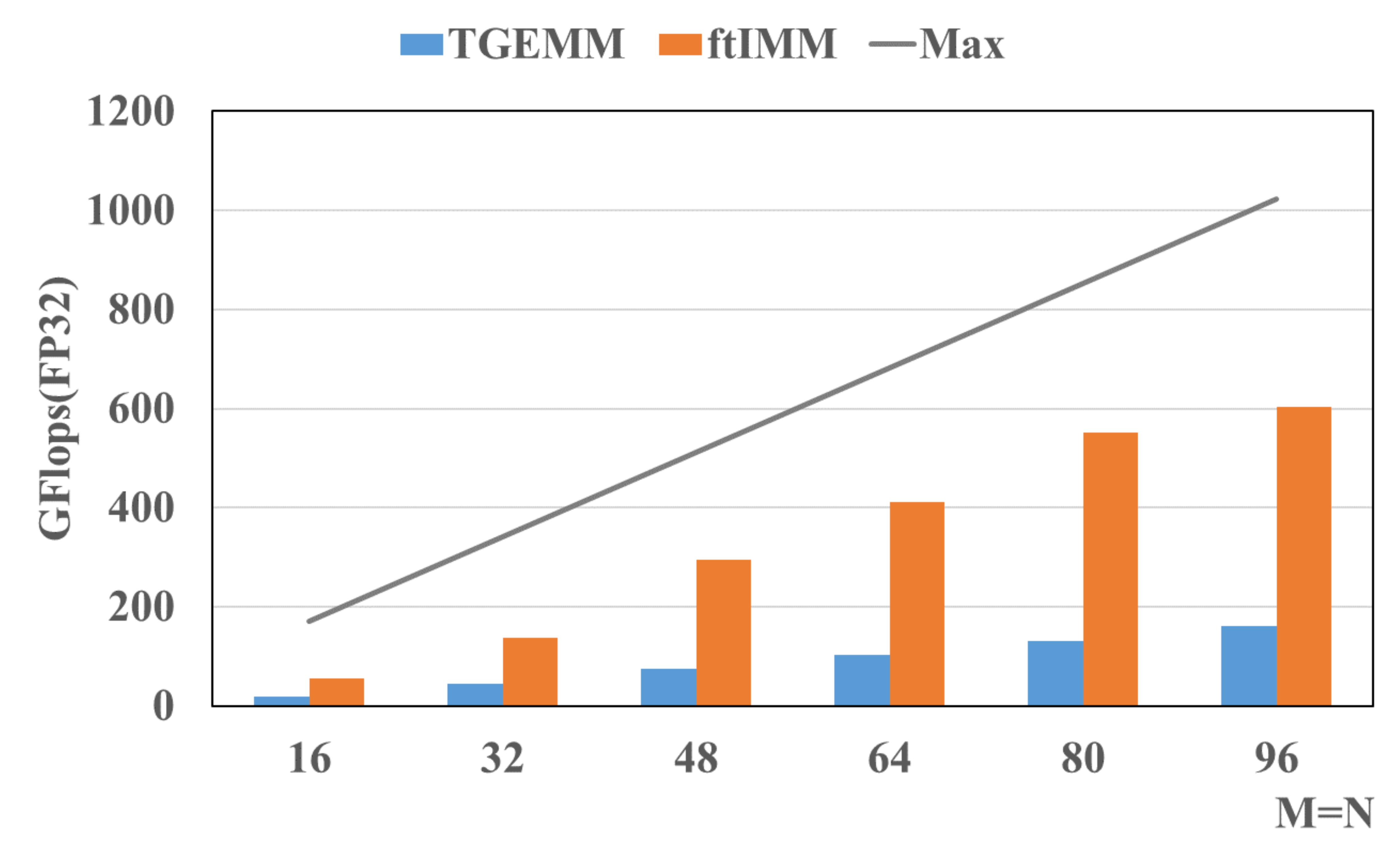}
            \end{minipage}%
            \label{fig:MultiCorePerformance-sf2}
    }
    \subfigure[$M=K=20480$]{
        \begin{minipage}[t]{0.3\linewidth}
            \centering
            \includegraphics[width =1\textwidth]{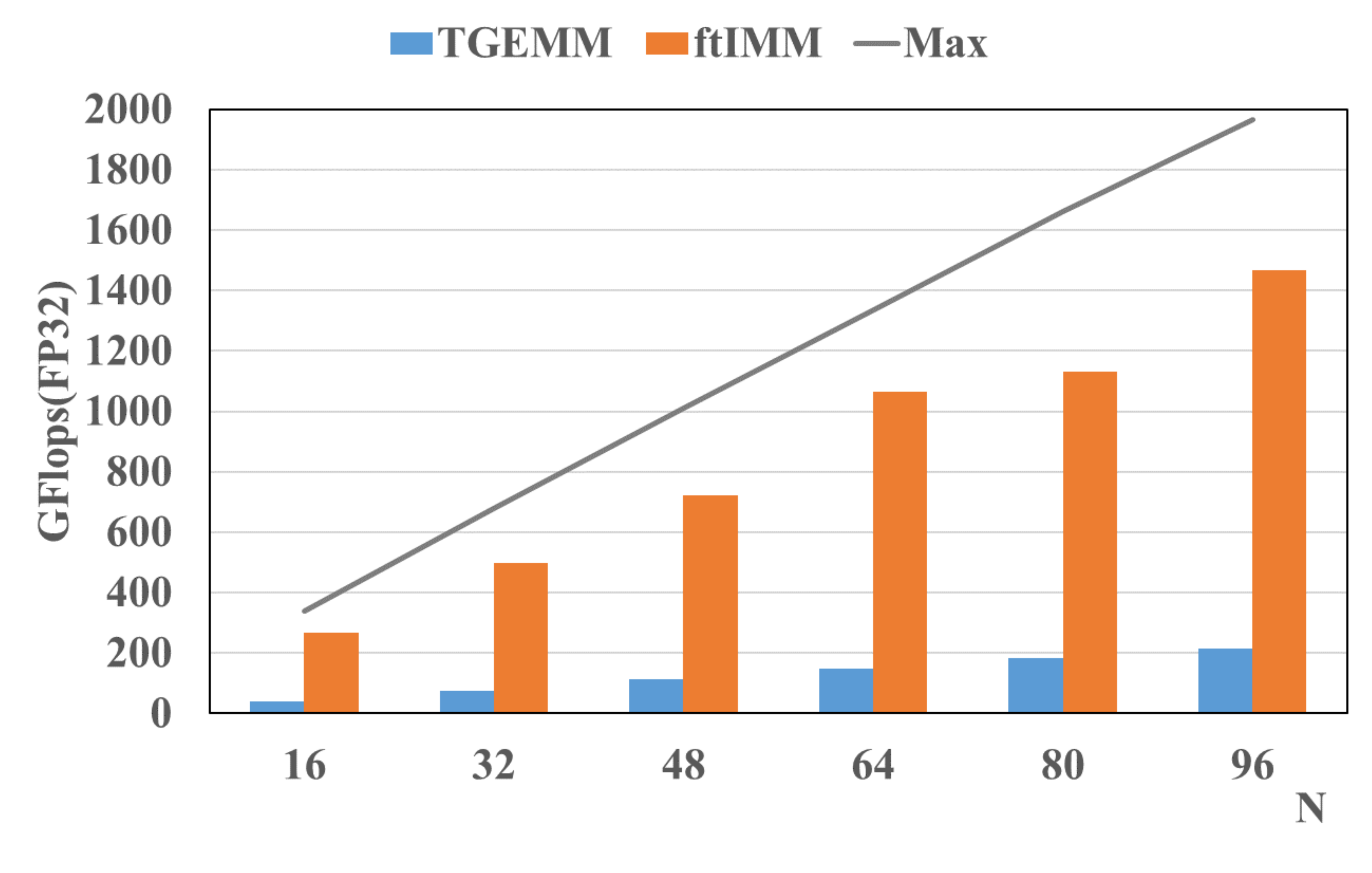}
            \end{minipage}%
            \label{fig:MultiCorePerformance-sf3}
    }
    
    \subfigure[$N=K=32$]{
        \begin{minipage}[t]{0.3\linewidth}
            \centering
            \includegraphics[width =1\textwidth]{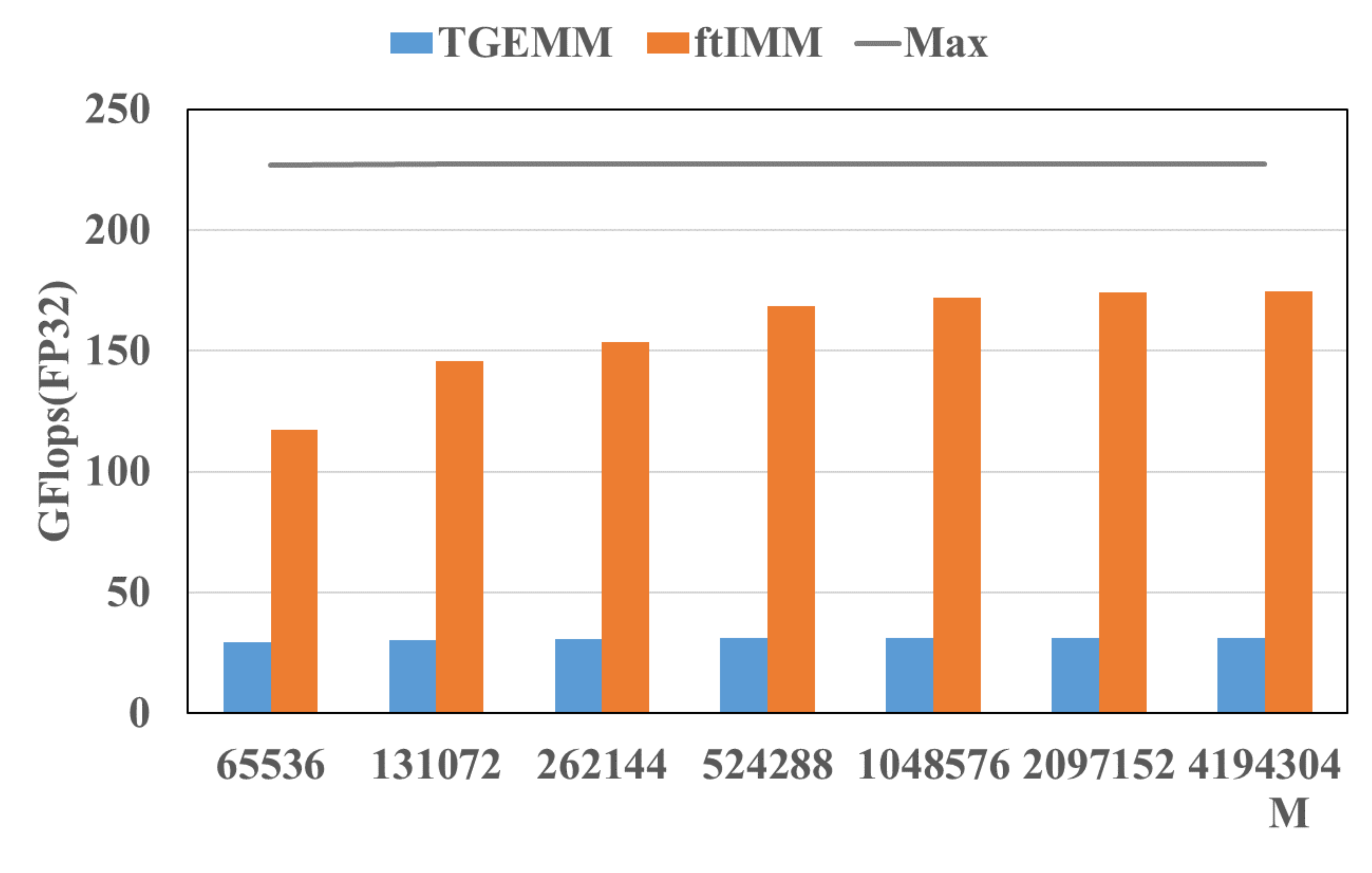}
            \end{minipage}%
            \label{fig:MultiCorePerformance-sf4}
    }%
    \subfigure[$M=N=32$]{
        \begin{minipage}[t]{0.3\linewidth}
            \centering
            \includegraphics[width =1\textwidth]{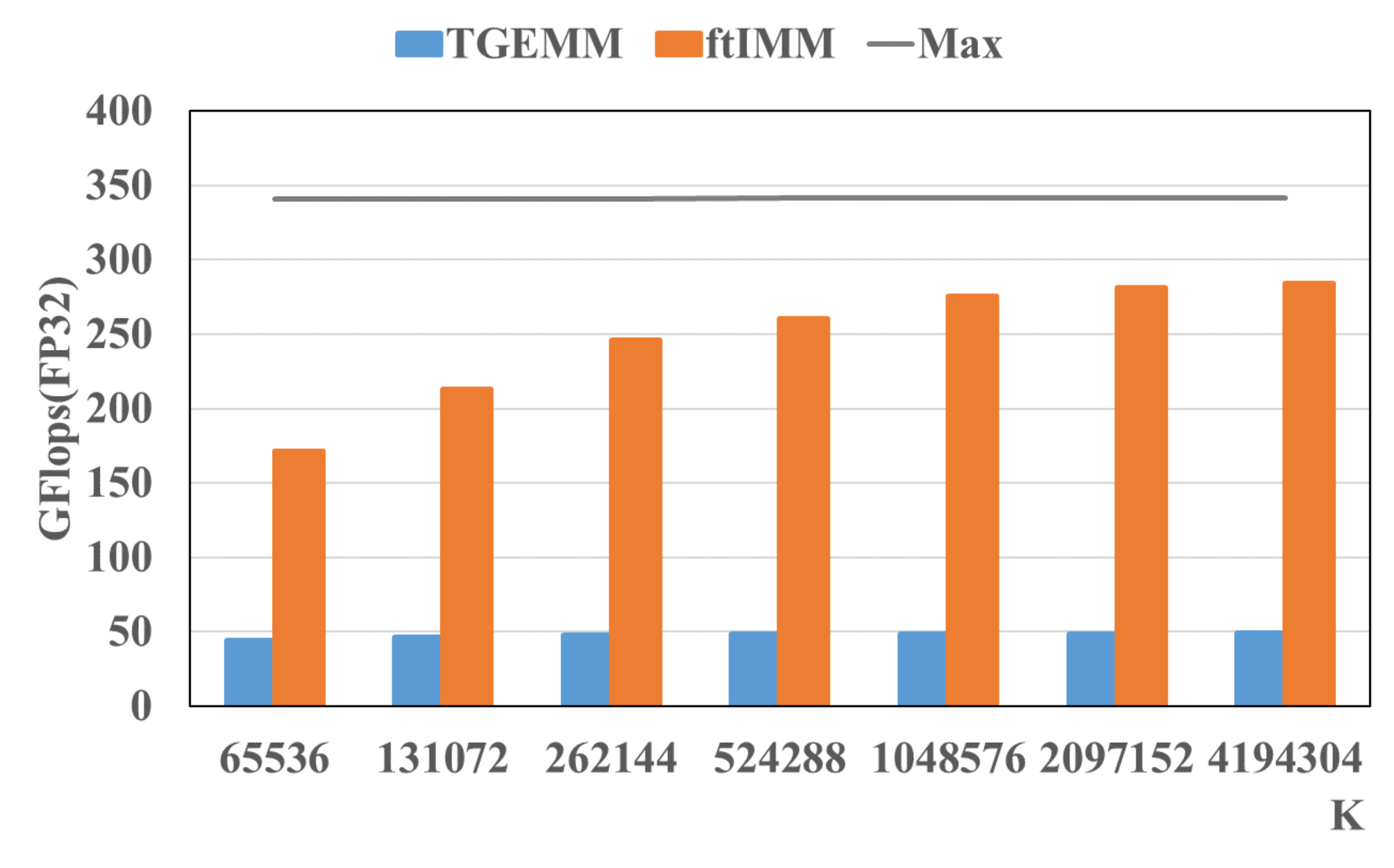}
            \end{minipage}%
            \label{fig:MultiCorePerformance-sf5}
    }
    \subfigure[$N=32$]{
        \begin{minipage}[t]{0.3\linewidth}
            \centering
            \includegraphics[width =1\textwidth]{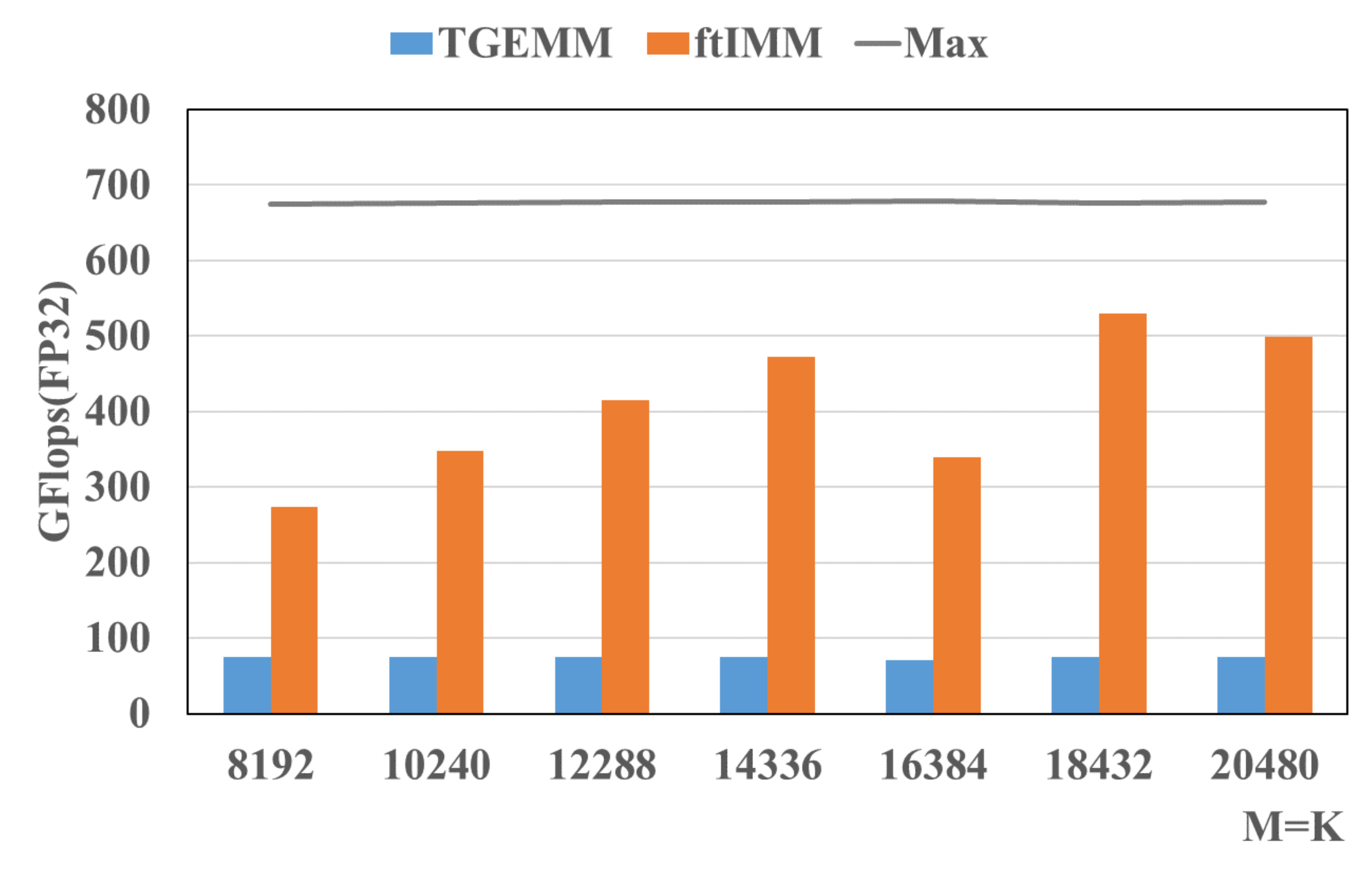}
            \end{minipage}%
            \label{fig:MultiCorePerformance-sf6}
    }
    \caption{Performance of ftIMM on Multiple Cores of a GPDSP Cluster}
    \label{fig:MultiCorePerformance}
\end{figure*}

% \begin{figure}[htb]
%     \centering
%     \includegraphics[width=\linewidth]{imgv4/scalabilty_MK_all.pdf}
%     % \includegraphics[width =1\textwidth]{imgv5_ftIMM/image029.pdf}
%     \caption{Scalability}
%     \label{fig:fig22}
% \end{figure}

\begin{figure}[htb]
    \centering
    \includegraphics[width=\linewidth]{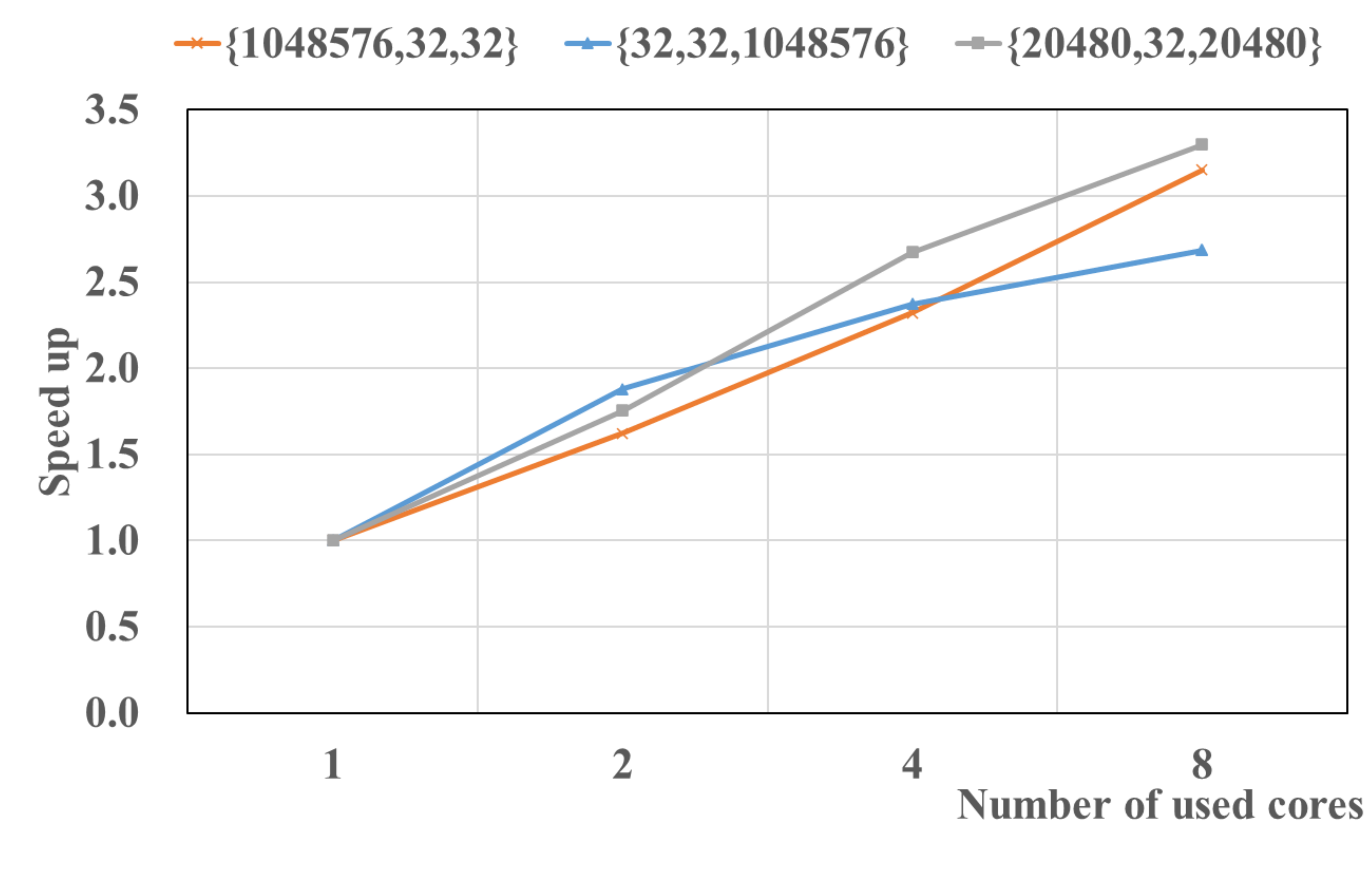}
    \caption{Scalability}
    \label{fig:fig22}
\end{figure}

\begin{figure*}[htb]
    \centering
    \subfigure[$M = 2^{16}$]{
        \begin{minipage}[t]{0.3\linewidth}
            \centering
            \includegraphics[width =1\textwidth]{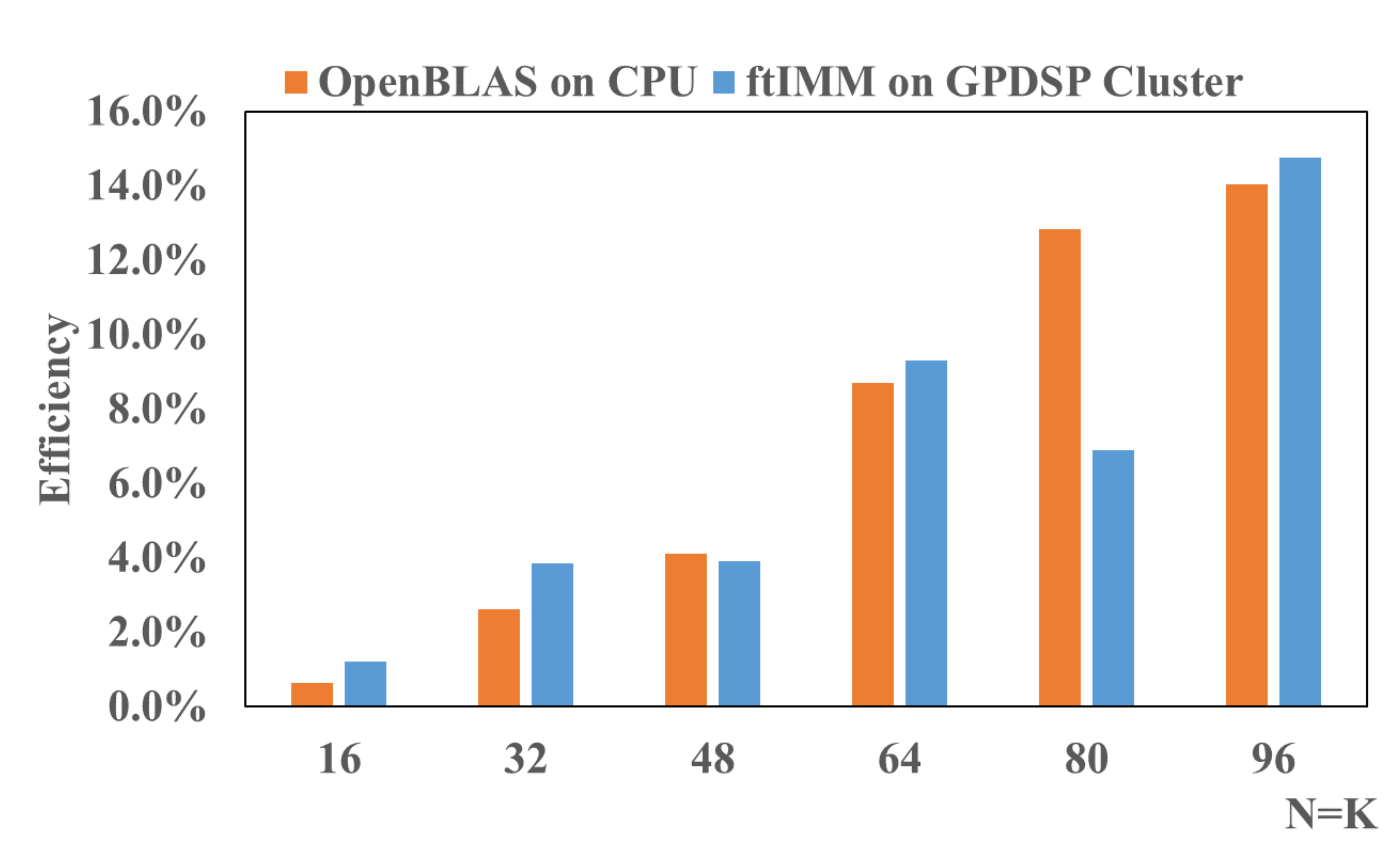}
            \end{minipage}%
    }
    \subfigure[$K = 2^{16}$]{
        \begin{minipage}[t]{0.3\linewidth}
            \centering
            \includegraphics[width =1\textwidth]{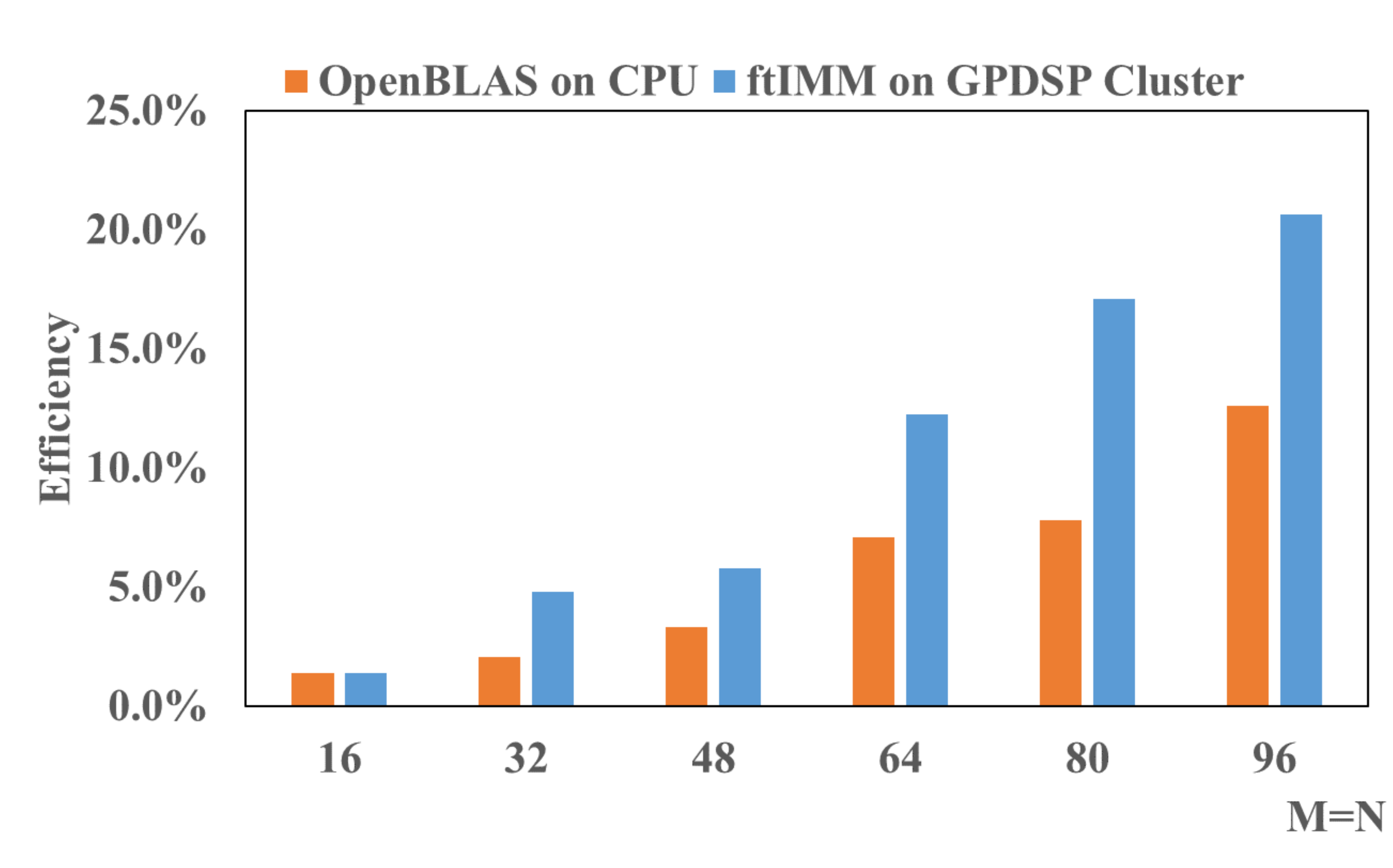}
            \end{minipage}%
    }
    \subfigure[$M = K = 20480$]{
        \begin{minipage}[t]{0.3\linewidth}
            \centering
            \includegraphics[width =1\textwidth]{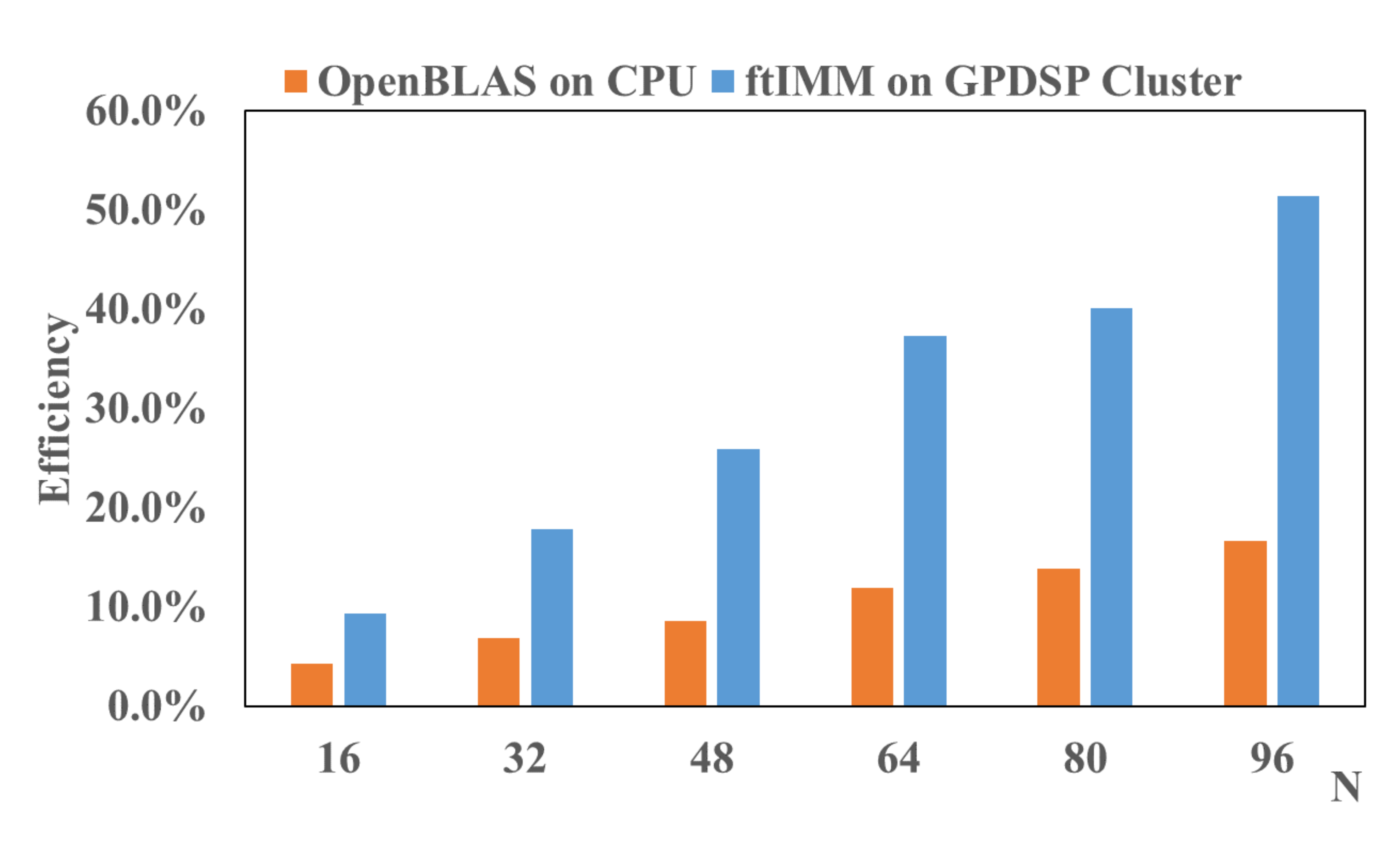}
            \end{minipage}%
    }
    \caption{Performance of Irregular-Shaped GEMM on CPU and GPDSP Cluster}
    \label{fig:CPU_DSP}
\end{figure*}

\subsection{Multi-core Performance}
We also evaluate ftIMM by applying it to the three types of irregular-shaped GEMMs using multiple cores of a GPDSP cluster on FT-m7032. Inputs and outputs are both single-precision matrices. The results are shown in Fig. \ref{fig:MultiCorePerformance}

\subsubsection{Multiplication between a tall-and-skinny matrix and a small matrix}
Fig. \ref{fig:MultiCorePerformance-sf1} and \subref{fig:MultiCorePerformance-sf4} show the results on multiplication between a tall-and-skinny matrix and a small matrix on a GPDSP cluster. 
In Fig. \ref{fig:MultiCorePerformance-sf1}, the size of the M dimension is set to $2^{16}$ which is large sufficiently. FtIMM achieves higher performance than TGEMM, yielding on up to $4.2\times$ performance improvement. The performance improvement is due to the parallelization strategy which takes full advantage of multiple cores, the optimized micro-kernels and dynamic adjusting. FtIMM also demonstrates higher performance benefit for smaller N and K sizes. This is largely due to the optimized micro-kernels and dynamic adjusting. The maximum performance of ftIMM obtained with the roofline model is also shown in Fig. \ref{fig:MultiCorePerformance}. FtIMM delivers up to $67.0\%$ of the maximum performance in the results. Because the performance of ftIMM on multiple cores is limited by the bandwidth and the actual bandwidth cannot reach the theoretical bandwidth which is used in the computation of roofline model, ftIMM cannot deliver maximum performance. 

% The implementation also demonstrates higher performance with larger N and K sizes. For example, in figure 11, for GEMMs with N = 96 and K = 96, the methods gives $4.3 \times$ higher GFLOPS over GEMMs with N = 32 and K = 32. This is mainly because with larger sizes of N and K dimensions, the algorithm can take full advantage of VPEs.

In Fig. \ref{fig:MultiCorePerformance-sf4}, the sizes of K and N dimensions are set to 32. We can observe that the performance benefit tends to be more significant for larger size of the M dimension. For example, for GEMMs with $M = 2^{22}$, ftImm gives higher improvement than GEMMs with M = $2^{16}$. This is mainly because larger size of the M dimension makes it easier for ftIMM to improve the reuse of data and to deliver higher performance.

\subsubsection{Multiplication between a skinny-and-tall matrix and a tall-and-skinny matrix}
Fig. \ref{fig:MultiCorePerformance-sf2} and \subref{fig:MultiCorePerformance-sf5} show the performance on the multiplication between a skinny-and-tall matrix and a tall-and-skinny matrix on a GPDSP cluster. In Fig. \ref{fig:MultiCorePerformance-sf2}, the size of K dimension is set to $2^{16}$. In Fig. \ref{fig:MultiCorePerformance-sf5}, the sizes of M and N dimensions are set to 32. The results demonstrate that for these cases, ftIMM delivers up to $5.8\times$ higher performance than TGEMM. Though the parallelization strategy chosen in these cases brings additional overhead of reduction, ftIMM can take full advantages of multiple cores especially when M and N are much smaller than K. We also observe that, because too small sizes of M and N dimensions cannot make full use of the computation ability and bandwidth, the performance is higher along with larger sizes of M and N dimensions.

\subsubsection{Multiplication between a large regular matrix and a tall-and-skinny matrix}
Fig. \ref{fig:MultiCorePerformance-sf3} and \subref{fig:MultiCorePerformance-sf6} show the results of the multiplication between a large regular matrix and a tall-and-skinny matrix on a GPDSP cluster. In Fig. \ref{fig:MultiCorePerformance-sf3}, the sizes of the M and K dimensions are set to $20480$. As can be seen from the results, ftIMM gives higher performance on this case, yielding on up to $7.2\times$ performance improvement. In Fig. \ref{fig:MultiCorePerformance-sf6}, the size of the N dimension is set to 32. It is can be seen from the results that ftIMM and TGEMM both achieve higher performance in the computation of the third type of irregular-shaped GEMMs than the other types. It is because in the third type of irregular-shaped GEMMs, the micro-kernel can achieve higher performance with larger size of K dimension than others and additional overhead can be amortized more sufficiently. For cases in which $M = K = 16384$ or $M = K = 20480$, ftIMM gives lower performance than other cases. This is because computation ability of used cores cannot be utilized sufficiently under the sizes of matrices and the corresponding block sizes.
%%%%%%%%%%%%%%%%%%%%%%%%%%%%%%%%%%%%%%%%%%%

\subsubsection{Scalability}
% Figure 13 and 14 show the scalability of the proposed implementations on performing the irregular-shaped GEMM of ${M \times N \times K}$ = ${2^{20} \times 32 \times 32}$ and ${32 \times 32 \times  2^{20}}$. It can be seen from the results that the proposed implementations, including the parallelization strategies in the M and K dimension, not only outperform the traditional implementation in the irregular-shaped GEMMs, but also show the best scalability as the number of used cores increases.

Fig. \ref{fig:fig22} shows the scalability of ftIMM on performing the three types of irregular-shaped GEMMs of $M \times N \times K$ = $2^{20} \times 32 \times 32$, $32 \times 32 \times  2^{20}$ and $20480 \times 32 \times  20480$. The vertical axis represents the speedup of ftIMM on multiple cores of a GPDSP cluster compared with on the single DSP core. It can be seen from the results that the performance of ftIMM increases with the number of DSP cores, while the scaling efficiency is not high. This is mainly because the algorithm is memory intensive and the performance is limited by the bandwidth when more DSP cores are used. Besides, among the three types of irregular-shaped GEMMs, for the case $M \times N \times K = 20480 \times 32 \times  20480$, the scalability of the proposed algorithm is worse than other cases. This is because the parallelization strategy based on K dimension is chosen for this case and the overhead of reduction among multiple cores increases with the increase of the number of cores.

% We observe that the scalability is not so high as the number of cores increases.

\subsection{Performance of Irregular-shaped GEMMs on CPU and GPDSP Cluster of FT-m7032}
Though the open-source library LibShalom is optimized for irregular-shaped GEMMs on ARMv8 CPU, the available version can not work on the case of $C \; + = A \times B$. Therefore we compare the performance of ftIMM on a GPDSP cluster of FT-m7032 with the OpenBLAS-0.3.20, which is highly optimized for GEMM, on the 16-core ARMv8 CPU of FT-m7032 based on the same bandwidth. We evaluate the performance on the three types of irregular-shaped GEMMs and the results are shown in Fig. \ref{fig:CPU_DSP}. We can observe that ftIMM delivers higher efficiency, which is the ratio of the achieved performance to the corresponding peak performance, on a GPDSP cluster than OpenBLAS on the multi-core CPU of FT-m7032 in most cases, yielding on up to $3.1\times$ improvement.

% \begin{figure}[h]
%     \centering
%     \includegraphics[width=\linewidth]{imgv4/CPU_DSP1.pdf}
%     \caption{Performance of irregular-shaped GEMM on CPU and DSP}
%     \label{fig:fig23}
% \end{figure}\begin{figure}[h]
%     \centering
%     \includegraphics[width=\linewidth]{imgv4/CPU_DSP2.pdf}
%     \caption{Performance of irregular-shaped GEMM on CPU and DSP}
%     \label{fig:fig24}
% \end{figure}\begin{figure}[h]
%     \centering
%     \includegraphics[width=\linewidth]{imgv4/CPU_DSP3.pdf}
%     \caption{Performance of irregular-shaped GEMM on CPU and DSP}
%     \label{fig:fig25}
% \end{figure}

% \input{conclusion}
% \input{conclusion}

\section{Conclusion}
We propose ftIMM, an efficient implementation for irregular-shaped matrix-matrix multiplication on multi-core DSPs of FT-m7032. We present an auto-generation scheme for building kernels of different sizes. Besides, we design the multi-core algorithm for irregular-shaped GEMMs with two parallelization strategies and the dynamic adjusting function. We evaluate ftIMM by applying it to irregular-shaped GEMMs on multi-core DSPs of FT-m7032 and compare it with the traditional implementation, which shows that ftIMM achieves higher performance for irregular-shaped GEMMs.

% \section*{Acknowledgment}
% The corresponding author is Qinglin Wang, and this work was supported by the National Natural Science Foundation of China under Grant No. 62002365.

\bibliographystyle{IEEEtran}
\bibliography{Ref-20220513}

\begin{thebibliography}{10}
\providecommand{\url}[1]{#1}
\csname url@samestyle\endcsname
\providecommand{\newblock}{\relax}
\providecommand{\bibinfo}[2]{#2}
\providecommand{\BIBentrySTDinterwordspacing}{\spaceskip=0pt\relax}
\providecommand{\BIBentryALTinterwordstretchfactor}{4}
\providecommand{\BIBentryALTinterwordspacing}{\spaceskip=\fontdimen2\font plus
\BIBentryALTinterwordstretchfactor\fontdimen3\font minus
  \fontdimen4\font\relax}
\providecommand{\BIBforeignlanguage}[2]{{%
\expandafter\ifx\csname l@#1\endcsname\relax
\typeout{** WARNING: IEEEtran.bst: No hyphenation pattern has been}%
\typeout{** loaded for the language `#1'. Using the pattern for}%
\typeout{** the default language instead.}%
\else
\language=\csname l@#1\endcsname
\fi
#2}}
\providecommand{\BIBdecl}{\relax}
\BIBdecl

\bibitem{openblas2012}
Z.~Xianyi, W.~Qian, and Z.~Yunquan, ``Model-driven level 3 blas performance
  optimization on loongson 3a processor,'' in \emph{2012 IEEE 18th
  International Conference on Parallel and Distributed Systems}, 2012, pp.
  684--691.

\bibitem{GotoBlas}
K.~Goto and R.~Van De~Geijn, ``High-performance implementation of the level-3
  blas,'' \emph{ACM Trans. Math. Softw.}, vol.~35, no.~1, Jul 2008.

\bibitem{BLIS}
F.~G. Van~Zee and R.~A. van~de Geijn, ``Blis: A framework for rapidly
  instantiating blas functionality,'' \emph{ACM Trans. Math. Softw.}, vol.~41,
  no.~3, Jun 2015.

\bibitem{heinecke2016libxsmm}
A.~Heinecke, G.~Henry, M.~Hutchinson, and H.~Pabst, ``Libxsmm: accelerating
  small matrix multiplications by runtime code generation,'' in \emph{SC'16:
  Proceedings of the International Conference for High Performance Computing,
  Networking, Storage and Analysis}.\hskip 1em plus 0.5em minus 0.4em\relax
  IEEE, 2016, pp. 981--991.

\bibitem{dhillon2004kernel}
I.~S. Dhillon, Y.~Guan, and B.~Kulis, ``Kernel k-means: spectral clustering and
  normalized cuts,'' in \emph{Proceedings of the tenth ACM SIGKDD international
  conference on Knowledge discovery and data mining}, 2004, pp. 551--556.

\bibitem{drake2012accelerated}
J.~Drake and G.~Hamerly, ``Accelerated k-means with adaptive distance bounds,''
  in \emph{5th NIPS workshop on optimization for machine learning}, vol.~8,
  2012.

\bibitem{hamerly2010making}
G.~Hamerly, ``Making k-means even faster,'' in \emph{Proceedings of the 2010
  SIAM international conference on data mining}.\hskip 1em plus 0.5em minus
  0.4em\relax SIAM, 2010, pp. 130--140.

\bibitem{rivera2021tsm2x}
C.~Rivera, J.~Chen, N.~Xiong, J.~Zhang, S.~L. Song, and D.~Tao, ``Tsm2x:
  High-performance tall-and-skinny matrix--matrix multiplication on gpus,''
  \emph{Journal of Parallel and Distributed Computing}, vol. 151, pp. 70--85,
  2021.

\bibitem{jia2014caffe}
Y.~Jia, E.~Shelhamer, J.~Donahue, S.~Karayev, J.~Long, R.~Girshick,
  S.~Guadarrama, and T.~Darrell, ``Caffe: Convolutional architecture for fast
  feature embedding,'' in \emph{Proceedings of the 22nd ACM International
  Conference on Multimedia}, ser. MM '14.\hskip 1em plus 0.5em minus
  0.4em\relax New York, NY, USA: Association for Computing Machinery, 2014, p.
  675–678.

\bibitem{ijcnnwang2019}
Q.~Wang, M.~Songzhu, J.~Liu, and C.~Gong, ``Parallel convolution algorithm
  using implicit matrix multiplication on multi-core cpus,'' in \emph{2019
  International Joint Conference on Neural Networks (IJCNN)}, 07 2019, pp.
  1--7.

\bibitem{vgg2014}
K.~Simonyan and A.~Zisserman, ``Very deep convolutional networks for
  large-scale image recognition,'' 2015.

\bibitem{resnet}
S.~Targ, D.~Almeida, and K.~Lyman, ``Resnet in resnet: Generalizing residual
  architectures,'' 2016.

\bibitem{chen2019tsm2}
J.~Chen, N.~Xiong, X.~Liang, D.~Tao, S.~Li, K.~Ouyang, K.~Zhao,
  N.~DeBardeleben, Q.~Guan, and Z.~Chen, ``Tsm2: optimizing tall-and-skinny
  matrix-matrix multiplication on gpus,'' in \emph{Proceedings of the ACM
  International Conference on Supercomputing}, 2019, pp. 106--116.

\bibitem{rivera2020ism2}
C.~Rivera, J.~Chen, N.~Xiong, S.~L. Song, and D.~Tao, ``Ism2: Optimizing
  irregular-shaped matrix-matrix multiplication on gpus,'' \emph{arXiv preprint
  arXiv:2002.03258}, 2020.

\bibitem{ernst2021performance}
D.~Ernst, G.~Hager, J.~Thies, and G.~Wellein, ``Performance engineering for
  real and complex tall \& skinny matrix multiplication kernels on gpus,''
  \emph{The International Journal of High Performance Computing Applications},
  vol.~35, no.~1, pp. 5--19, 2021.

\bibitem{tang2021efficient}
H.~Tang, K.~Komatsu, M.~Sato, and H.~Kobayashi, ``Efficient mixed-precision
  tall-and-skinny matrix-matrix multiplication for gpus,'' \emph{International
  Journal of Networking and Computing}, vol.~11, no.~2, pp. 267--282, 2021.

\bibitem{yang2021libshalom}
W.~Yang, J.~Fang, D.~Dong, X.~Su, and Z.~Wang, ``Libshalom: optimizing small
  and irregular-shaped matrix multiplications on armv8 multi-cores,'' in
  \emph{Proceedings of the International Conference for High Performance
  Computing, Networking, Storage and Analysis}, 2021, pp. 1--14.

\bibitem{Li2021AutoTSMM}
C.~Li, H.~Jia, H.~Cao, J.~Yao, B.~Shi, C.~Xiang, J.~Sun, P.~Lu, and Y.~Zhang,
  ``Autotsmm: An auto-tuning framework for building high-performance
  tall-and-skinny matrix-matrix multiplication on cpus,'' in \emph{2021 IEEE
  Intl Conf on Parallel Distributed Processing with Applications, Big Data
  Cloud Computing, Sustainable Computing Communications, Social Computing
  Networking (ISPA/BDCloud/SocialCom/SustainCom)}, 2021, pp. 159--166.

\bibitem{igual2012unleashing}
F.~D. Igual, M.~Ali, A.~Friedmann, E.~Stotzer, T.~Wentz, and R.~A. van~de
  Geijn, ``Unleashing the high-performance and low-power of multi-core dsps for
  general-purpose hpc,'' in \emph{SC'12: Proceedings of the International
  Conference on High Performance Computing, Networking, Storage and
  Analysis}.\hskip 1em plus 0.5em minus 0.4em\relax IEEE, 2012, pp. 1--11.

\bibitem{tiwari2018high}
A.~Tiwari, V.~Kumar, and G.~Mitra, ``High performance and energy optimal
  parallel programming on cpu and dsp based mpsoc,'' Ph.D. dissertation,
  IIIT-Delhi, 2018.

\bibitem{wang2021advancing}
Y.~Wang, C.~Li, C.~Liu, S.~Liu, Y.~Lei, J.~Zhang, Y.~Zhang, and Y.~Guo,
  ``Advancing dsp into hpc, ai, and beyond: challenges, mechanisms, and future
  directions,'' \emph{CCF Transactions on High Performance Computing}, vol.~3,
  no.~1, pp. 114--125, 2021.

\bibitem{van2016blis}
F.~G. Van~Zee, T.~M. Smith, B.~Marker, T.~M. Low, R.~A. V.~D. Geijn, F.~D.
  Igual, M.~Smelyanskiy, X.~Zhang, M.~Kistler, V.~Austel \emph{et~al.}, ``The
  blis framework: Experiments in portability,'' \emph{ACM Transactions on
  Mathematical Software (TOMS)}, vol.~42, no.~2, pp. 1--19, 2016.

\bibitem{maDRAM2019}
S.~Ma, Z.~Liu, S.~Chen, L.~Huang, Y.~Guo, Z.~Wang, and M.~Zhang, ``Coordinated
  dma: Improving the dram access efficiency for matrix multiplication,''
  \emph{IEEE Transactions on Parallel and Distributed Systems}, vol.~30,
  no.~10, pp. 2148--2164, 2019.

\bibitem{liu2018matrix}
Z.~Liu and X.~Tian, ``\BIBforeignlanguage{Chinese}{Vectorization of matrix
  multiplication for multi-core vector processors},''
  \emph{\BIBforeignlanguage{Chinese}{Jisuanji Xuebao/Chinese Journal of
  Computers}}, vol.~41, no.~10, pp. 2251 -- 2264, 2018.

\bibitem{HUANG2021102248}
X.~Huang, Q.~Wang, S.~Lu, R.~Hao, S.~Mei, and J.~Liu, ``Evaluating fft-based
  algorithms for strided convolutions on armv8 architectures,''
  \emph{Performance Evaluation}, vol. 152, p. 102248, 2021.

\bibitem{huang2021numa}
------, ``Numa-aware fft-based convolution on armv8 many-core cpus,'' in
  \emph{2021 IEEE Intl Conf on Parallel \& Distributed Processing with
  Applications, Big Data \& Cloud Computing, Sustainable Computing \&
  Communications, Social Computing \& Networking
  (ISPA/BDCloud/SocialCom/SustainCom)}.\hskip 1em plus 0.5em minus 0.4em\relax
  IEEE, 2021, pp. 1019--1026.

\end{thebibliography}

\end{document}